**Title:** **The Most Difference in Means: A Statistic for the Strength of Null and Near-Zero Results**

**Authors:** Bruce A. Corliss[1,*], Taylor R. Brown[2], Tingting Zhang[3], Kevin A. Janes[4], Heman Shakeri[1], Philip E. Bourne[1,4]

**Affiliations**

[1]School of Data Science, University of Virginia; Charlottesville, Virginia

[2]Department of Statistics, University of Virginia; Charlottesville, Virginia

[3]Department of Statistics, University of Pittsburgh; Pittsburgh, Pennsylvania

[4]Department of Biomedical Engineering, University of Virginia; Charlottesville, Virginia

*Corresponding author.

**Contact Information**: Bruce Corliss, bac7wj@virginia.edu.

**Classification**: Statistics, Applied Biological Sciences

**Summary:**

Statistical insignificance does not suggest the absence of effect, yet scientists must often use null results as evidence of negligible (near-zero) effect size to falsify scientific hypotheses. Doing so must assess a result's null strength, defined as the evidence for a negligible effect size. Such an assessment would differentiate strong null results that suggest a negligible effect size from weak null results that suggest a broad range of potential effect sizes. We propose the most difference in means ($\delta_M$) as a two-sample statistic that can both quantify null strength and perform a hypothesis test for negligible effect size. To facilitate consensus when interpreting results, our statistic allows scientists to conclude that a result has negligible effect size using different thresholds with no recalculation required. To assist with selecting a threshold, $\delta_M$ can also compare null strength between related results. Both $\delta_M$ and the relative form of $\delta_M$ outperform other candidate statistics in comparing null strength. We compile broadly related results and use the relative $\delta_M$ to compare null strength across different treatments, measurement methods, and experiment models. Reporting the relative $\delta_M$ may provide a technical solution to the file drawer problem by encouraging the publication of null and near-zero results.

**Introduction:**

Two-sample p-values from null hypothesis significance tests remain the gold standard for the analysis and reporting of scientific results despite calls to discontinue or de-emphasize their use (*1*, *2*). P-values can differentiate positive results (statistically significant) from null results (statistically insignificant). Yet p-values cannot give any indication of the practical equivalence of results: whether the observed effect size is close enough to zero to be considered negligible. Practically equivalent results play a key role in scientific research by falsifying scientific hypotheses and offering contrary evidence to previously reported positive results (*3*). Such an assessment would differentiate strong null results that suggest a range of all negligible effect sizes from weak null results that suggest a range that includes non-negligible effect sizes.

Characterizing practical equivalence requires assessing the data, its context, and the perspectives of scientists. An effective statistic should be useful for all these tasks. A practically equivalent result has a negligible effect size, which requires some form of hypothesis test to determine if the effect size is less than a maximum threshold for what is negligible. Selecting the value of this threshold is context-specific and can greatly differ between scientists with differing perspectives. Scientists with different perspectives will select different values for this threshold, yet collectively need to reach consensus for which results are practically equivalent. A useful statistic should allow each scientist to test for practical equivalence according to their threshold without having to reanalyze the data. Selecting an appropriate threshold is a critical part of this analysis and reviewing the null strength of related results should inform this selection. A useful statistic should also facilitate threshold selection by allowing for the comparison of null strength between results and highlight noteworthy results that have exceptionally strong practical equivalence. To simplify the process for data analysis, it would be ideal to use a single statistic for these tasks. A statistic that can be used for multiple tasks gives a more useful and informative summary because the data can be interpreted in different contexts.

We present the most difference in means ($\delta_M$) as a statistic that is capable of all these tasks. This statistic allows hypothesis testing for negligible effect against any threshold value. Null strength can be compared between related results to inform the selection of a threshold and highlight results with exceptionally strong null strength. All of this can be done without recalculation of the statistic. To test our statistic against previously developed candidates, we characterize the multidimensional problem of assessing null strength with various functions of population parameters. These functions serve as ground truth for simulation testing. We use an integrated risk assessment to test $\delta_M$ and the relative form of $\delta_M$ ($r\delta_M$) against several candidate statistics by evaluating their error rates in comparing the null strength between simulated experiment results. Our statistics were the only candidates that demonstrated better than random error rates across all investigations. We illustrate with real data how $r\delta_M$ can be used to test for negligibility and compare the null strength of results from broadly related experiments that have a combination of different experiment models, conditions, populations, species, timepoints, treatments, and measurement techniques. We propose that reporting $r\delta_M$ of null and near-zero results will provide a more useful interpretation than alternative analysis techniques.

**Background**

***Bayesian Summary of Difference in Means***

Let $X_1, ..., X_m$ be an i.i.d. sample from a control group with a distribution *Normal*($\mu_X$, $\sigma_X^2$), and $Y_1, ..., Y_n$ be an i.i.d. sample from an experiment group with a distribution *Normal*($\mu_Y$, $\sigma_Y^2$). Both samples are independent from one another, and we conservatively assume unequal variance, i.e., $\sigma_X^2 \neq \sigma_Y^2$ (the Behrens-Fisher problem (*4*) for the means of normal distributions).

We analyze data in a Bayesian manner using minimal assumptions and therefore use a noninformative prior (*5*), specified as

$$p(\mu_X, \mu_Y, \sigma_X^2, \sigma_Y^2) \propto (\sigma_X^2)^{-1} (\sigma_Y^2)^{-1}. \tag{1}$$

The model has a closed-form posterior distribution. Specifically, the population means, conditional on the variance parameters and the data, follow normal distributions:

$$\mu_X \,|\, \sigma^2, x_{1:m} \sim Normal\left(\bar{x}, \frac{\sigma_X^2}{m}\right) \text{ and} \tag{2}$$

$$\mu_Y \,|\, \sigma^2, y_{1:n} \sim Normal\left(\bar{y}, \frac{\sigma_Y^2}{n}\right). \tag{3}$$

Moreover, the population variances each independently follow an inverse gamma distribution (*InvGamma*):

$$\sigma_X^2 \,|\, x_{1:m} \sim InvGamma\left(\frac{m-1}{2}, \frac{(m-1)s_X^2}{2}\right) \text{ and} \tag{4}$$

$$\sigma_Y^2 \,|\, y_{1:n} \sim InvGamma\left(\frac{n-1}{2}, \frac{(n-1)s_Y^2}{2}\right). \tag{5}$$

We exclude the use of prior information in this analysis because we wish to summarize the data alone and not be influenced by the beliefs of the scientist analyzing the data (specifically, the strength of the prior used can considerably influence the outputs of a Bayesian statistical analysis (*6, 7*)).

### Practical Equivalence Over a Raw Scale

We define that there is *stronger raw practical equivalence* when the absolute difference in population means between groups ($|\mu_{DM}|$) is smaller, where

$$|\mu_{DM}| = |\mu_Y - \mu_X|. \tag{6}$$

We summarize the posterior distribution of $|\mu_{DM}|$ to convey the evidence of raw practical equivalence from sample data. For the sake of simplicity, we report a single number. We define *raw null strength* as a conservative estimate of how large $|\mu_{DM}|$ could be based on sample data and a given credible level. Higher raw null strength (lower values of this estimate) conveys stronger evidence of raw practical equivalence.

Our proposed statistic to quantify raw null strength is the *most difference in means* ($\delta_M$), defined as an upper quantile of the posterior of $|\mu_{DM}|$. Specifically, if $Q_{raw}(p)$ is the quantile function of this posterior, then $\delta_M$ satisfies

$$\delta_M = Q_{raw}(1 - \alpha_{DM}) \tag{7}$$

for some "confidence" or credible level *1-a_{DM}*.

However, this quantile function is difficult to compute, and there is no closed-form posterior distribution for the transformed quantity $|\mu_Y - \mu_X|$. We estimate its distribution using the Monte Carlo method (*8*), which simulates values from the posterior of the untransformed mean parameters.

We estimate the population absolute difference in means using an upper quantile of an empirical cumulative distribution function (ECDF). We do this by exploiting the fact that $-c < \mu_Y - \mu_X < c$ if and only if $|\mu_Y - \mu_X| < c$, and we define $F_{raw}(x)$ as the ECDF of the signed difference in sample means from $K$ Monte Carlo simulations. With $K$ samples from the posterior distribution of $\mu_y$ and $u_x$, defined as the product of

$$\mu_X | x_{1:m} \sim t_{m-1}(\bar{x}, s_x^2 / m) \ \ and \tag{8}$$

$$\mu_y | y_{1:n} \sim t_{n-1}(\bar{y}, s_y^2 / n), \tag{9}$$

the cumulative distribution function is defined as

$$F_{raw}(x) = K^{-1} \sum_{i=1}^{K} \mathbb{I}(\mu_Y^i - \mu_X^i \leq x). \tag{10}$$

Then we numerically solve for the value of $c$ such that

$$F_{raw}(c) - F_{raw}(-c) = 1 - \alpha_{DM}. \tag{11}$$

Because this interval is centered at 0 (i.e. (-c,c)), we can report only the upper tail.

As defined above, $\delta_M$ is associated with a percentage $(1 - \alpha_{DM})$ to specify the credibility level in the same way that credibility intervals are annotated (i.e., 95% $\delta_M$ has a credibility level of 0.95). Indeed, simulations of the posterior distribution confirms that credibility of the $\delta_M$ approximates (1 - $\alpha_{DM}$) over a range of values (Fig. S1 A-C). Colloquially, the value of $\delta_M$ represents the largest absolute difference between the population means of the experiment group and control group supported by the data (visualized in Fig. 1A). Lower values of $\delta_M$ convey higher null strength between two groups' population means and suggest stronger practical equivalence.

***Practical Equivalence Over a Relative Scale***

To compare the practical equivalence of results across loosely related experiments, we extend the concept of raw practical equivalence to a relative scale. We define that there is stronger *relative practical equivalence* when the absolute relative difference between population means ($|r\mu_{DM}|$) is smaller (assuming $\mu_X > 0$), where

$$|r\mu_{DM}| = \left|\frac{\mu_Y - \mu_X}{\mu_X}\right|. \tag{12}$$

We define *relative null strength* as a conservative point estimate of how large $|r\mu_{DM}|$ could be based on sample data and a specified credibility level. Results with higher relative null strength (lower values of estimate) convey stronger evidence of relative practical equivalence. To calculate this estimate, we again begin by estimating a credible interval that has a (1-$\alpha_{DM}$) % probability to contain $|r\mu_{DM}|$. Just as we did above, we force the left endpoint of this interval to be 0. By reporting the upper bound of this credibility interval, we conservatively assess the range of likely values for $|r\mu_{DM}|$ from sample data.

While there is no closed-form posterior distribution for $|r\mu_{DM}|$ either, we can estimate its upper quantile using Monte Carlo simulations of the same posterior distribution derived from the prior and likelihood introduced in definitions in Eq. (8) – (9). We exploit the fact that $-c < r\mu_{DM} < c$ if and only if $|r\mu_{DM}| < c$, and we define $F_{relative}(x)$ as the ECDF of the signed relative difference in means approximated with Monte Carlo simulations from the posterior:

$$F_{relative}(x) = K^{-1} \sum_{i=1}^{K} \mathbb{I}\left(\frac{\mu_Y^i - \mu_X^i}{\mu_X^i} \leq x\right). \tag{13}$$

We define the *relative most difference in means* (r$\delta_M$) as the one-tailed upper quantile of |r$\mu_{DM}$|. As before, we numerically solve for the value of $c$ such that

$$F_{relative}(c) - F_{relative}(-c) = 1 - \alpha_{DM}. \tag{14}$$

Because this interval is centered at 0 (i.e. (-c,c)), we report only the upper tail. This upper bound is approximately equal to $Q_{relative}(p)$ with p set to the credible level $\alpha_{DM}$:

$$r\delta_M = Q_{relative}(1 - \alpha_{DM}). \tag{15}$$

The r$\delta_M$ is associated with a percentage (1 - $\alpha_{DM}$) to denote the credibility level as with $\delta_M$. Indeed, simulations of the posterior distribution confirms that credibility of r$\delta_M$ is equal to (1 - $\alpha_{DM}$) (Fig. S1 D-F). Colloquially, the value of r$\delta_M$ represents the largest absolute percent difference between the population means of the experiment group and control group supported by the data. Results with lower values of r$\delta_M$ have lower relative null strength and suggest stronger relative practical equivalence.

### Hypothesis Testing for Negligibility with $\delta_M$ and $r\delta_M$

To determine if a result is practically equivalent, scientists can perform a hypothesis test by testing if the magnitude of $\mu_{DM}$ and r$\mu_{DM}$ is less than a chosen threshold. Our statistics are a form of equal-tailed credible intervals, which are typically used for parameter estimation rather than hypothesis testing. However, credible intervals have been used for hypothesis tests against any threshold at the same credibility level as the interval (*9–12*). The procedure checks if the threshold is within the bounds of the interval. In this sense, intervals can be used for hypothesis testing against any threshold without recalculation of the interval. We note that there is controversy with the use of credibility intervals for hypothesis testing because the size of the effect is estimated under the assumption that it exists (*13*). We perceive this reservation to be a nonissue because an effect-size of zero has never been shown to exist in the real world and can't be confirmed with finite data (*14*). We assume a non-zero effect size is present in all cases, the question this procedure asks is whether there is evidence that it is small enough to be considered negligible.

We perform a hypothesis test against the specified threshold δ of the form

$$H_0: |\mu_{DM}| \geq \delta; \ H_1: |\mu_{DM}| < \delta. \tag{16}$$

We reject $H_0$ and conclude practical equivalence if $\delta_M < \delta$ because $\delta_M$ is the upper bound of the posterior for $|\mu_{DM}|$.

We illustrate the hypothesis testing procedure with a collection of hypothetical results from related experiments that measure the same phenomenon (Fig. 1B-E). For this example, two scientists choose different thresholds for negligible effect size (δ, δ'), test for negligible effect based on their own threshold, and then arrive at a consensus for which results are practically equivalent. These results are summarized by reporting the value for $\delta_M$ (Fig. 1B) approximated as the maximum of the absolute credible bounds of $\mu_{DM}$ (approximation used to simplify visualization, see Fig. S2). The first scientist uses a threshold of δ and performs a hypothesis test (Eq. 16) for each of these results. The results that reject the null hypothesis are designated as practically equivalent (Fig 2C, rows a, b, d, e, f, g) without any recalculation of $\delta_M$ required. Visually, this hypothesis test is equivalent to checking that the credible interval falls within the null region $H_0^\pm$ defined by [-δ, +δ] and closely resembles the procedure used in second generation p-values in concluding full support for the null hypothesis with a null interval centered about zero (*15*). If the credible interval overlaps with the inside and outside of the null region, then the expression $\delta_M \geq \delta$ is true, and the result is designated as not practically equivalent (Fig 2C, rows c, h, j). Similarly, the same designation occurs when the credible interval falls completely outside of the null region (Fig 2C, rows i, k). Meanwhile, a second scientist chooses a 30% smaller threshold for negligibility (δ'). The second scientist performs the same hypothesis testing procedure with δ' and designates practical equivalence for any result with a $\delta_M$ within the null region $H'_0{}^\pm$ (Fig 2D, rows a, b, d, e). These scientists can reach a consensus for which results are practically equivalent by identifying instances where they make the same designation (Fig 2E, rows a, b, d, e). The same procedure can be performed with $r\delta_M$ using relative units for the threshold and intervals, along with including results from experiments that are more broadly related (see Applied Examples section).

It is important to note that the first and second scientist may represent two scientists in the same field with differing opinions about what is negligible, or from different fields that have different requirements for negligibility. For instance, the threshold for negligibility for a measurement of an adverse side-effect may be larger (more forgiving) for an acute treatment versus long-term, or for an adult treatment versus pediatric. Indeed, the second scientist may even represent the first scientist in the future when their expectation for negligible effect size is more stringent. For example, an acceptable intensity of an adverse side-effect would be more forgiving for the first-in-class treatment of a particular disease versus after decades of development when several competing alternative treatments are available.

### *Measures of Raw and Relative Null Strength*

While we have proposed two statistics to quantify the evidence of practical equivalence, we need to develop a structured characterization of null strength to assess efficacy. This assessment relies on identifying the parameters that alter null strength on a raw and relative scale. Null strength is difficult to characterize in a controlled fashion because it depends on several parameters in addition to $\mu_X$ and $\mu_Y$. To characterize our statistics in a controlled fashion, we decompose null strength into a set of previously defined functions of population parameters. These functions are used as measures of null strength. We can vary each of these in isolation and study the effects they produce on our statistic.

For assessing raw null strength between population means, we identify a set of four measures that can be altered independently ($|\mu_{DM}|$, $\sigma_D$, $df_D$, and $\alpha_{DM}$ defined in Table 1, Fig. 2 A-E, see Materials and Methods: Explanation of Raw Null Strength Measures). For assessing the relative null strength between population means, we divide the same null strength measures by the control group mean when appropriate to form another set of variables that can be altered independently ($|r\mu_{DM}|$, $r\sigma_D$, $df_D$, and $\alpha_{DM}$ defined in Table 1, Fig. 2 G-K, see Materials and Methods: Explanation of Relative Null Strength Measures). Note that some relative null strength measures cannot be altered independently from raw null strength measures (e.g., altering $|\mu_{DM}|$ can also change $|r\mu_{DM}|$ or $r\sigma_D$).

## Results

### $\delta_M$ and $r\delta_M$ Covary with Changes to Null Strength

A statistic that effectively estimates raw null strength should covary with each measure of raw null strength in a consistent direction. We generated a series of population parameter configurations where each measure of raw null strength was individually altered towards higher raw null strength (stronger evidence of raw practical equivalence). The mean of various candidate statistics was computed on repeated samples drawn from these configurations (candidate statistics listed in Table S1). The mean of a useful statistic should either increase for all null strength measures or decrease. We found that the mean values from equivalence p-values and $\delta_M$ had a significant rank correlation in a consistent direction with null strength for all measures (Fig. 2F, Fig S3-S4, null region interval set to [-1, 1] from control sample mean for BF, $P_E$, and $P_\delta$). Additionally, we generated sets of population parameter configurations where each measure of relative null strength was altered towards higher relative null strength. The mean values from equivalence p-values and $r\delta_M$ had a significant rank correlation in a consistent direction for all measures (Fig. 2L, Fig. S5-S6, null region interval set to [-10%, 10%] of control sample mean for BF, $P_E$, and $P_\delta$). This initial analysis had potential confounding effects since $\mu_{DM}$ and $r\mu_{DM}$ could not be altered independently from the other measures. It is also important to note that we are testing candidate statistics for tasks that they were not designed for.

### $\delta_M$ *and* $r\delta_M$ *Exhibit Lower Comparison Error than Candidate Statistics*

We next performed a risk assessment to examine how effective the candidate statistics were at determining which of two results had higher null strength and deemed more noteworthy (see Supplementary Methods: Integrated Risk Assessment of Null Strength). We represented this decision of higher null strength with a 0-1 loss function that determined if the candidate statistics' prediction of higher null strength with the ground truth established by each measure of null strength. For a single population configuration, we calculated the expected value of the loss function to assess frequentist risk (*16*). This frequentist risk is the comparison error, defined as the probability of making an incorrect prediction for higher null strength. To explore general trends across the parameter space, we averaged comparison errors from many different parameter configurations with a similar approach to calculating various forms of integrated risk (*16*). Population configurations were stratified based on the expected t-ratio, approximated by Monte Carlo samples. The expected t-ratio is defined as the mean t-statistic of $\mu_{DM}$ across samples scaled to the critical value (denoted as $\bar{t}_{statistic}$ / $|t_{critical}|$, see Supplementary Materials and Methods: Parameter Space for Population Configurations). Population configurations were separated between those associated with null results (absolute expected t-ratio ≤ 1) and critical results (absolute expected t-ratio > 1). Investigations for comparison error were conducted for each of the four independent measures for null strength, both individually and simultaneously. $\delta_M$ was the only candidate statistic that exhibited a comparison error rate lower than random 50/50 guessing for all simulation studies for raw null strength (Fig. 3A, Fig. S7-S10, null region interval set to [-1, 1] from sample mean for BF, $P_E$, and $P_\delta$). Similarly, $r\delta_M$ was the only candidate statistic that exhibited a comparison error rate lower than random for all simulation studies for relative null strength (Fig 3B, Fig. S11-S14, null region interval set to [-10%, 10%] of control sample mean for BF, $P_E$, and $P_\delta$).

***Applied Examples***

We compiled results from studies of atherosclerosis to illustrate how the $r\delta_M$ could be used to assess the null strength and noteworthiness of results. Atherosclerosis is the underlying cause of approximately 50% of all deaths in developed nations (*17*) and is characterized by the build-up of fatty deposits, called plaques, on the inner wall of arteries. Researchers use dietary, behavioral, pharmacological, and genetic interventions to study atherosclerosis and measure various biological phenomenon to monitor disease severity, including plasma cholesterol and plaque size.

While lowering total plasma cholesterol is therapeutic in most cases (depending on the composition of the cholesterol (*17*)), many interventions treat atherosclerosis through other means. It is important to determine whether an intervention has a negligible effect size on total plasma cholesterol to help elucidate its underlying mechanisms. Plasma cholesterol levels vary from 60-3000 mg/dL across animal models used to research atherosclerosis and are reported in units of mmol/L as well (Table S3, S4). This large variation in the measurement values makes it necessary to evaluate null strength on a relative scale. The $r\delta_M$ is the only statistic that can be used to simultaneously test for negligibility using different thresholds and compare the null strength without recalculation (Fig. 4A-B). A literature search for positive results associated with reducing total cholesterol reveals that the relative difference in means is larger than 30% for most cases (Table S4). If this threshold is used to delineate a maximum negligible effect size, these null results could be separated based on a hypothesis test that designates them as practically equivalent ($r\delta_M < 30\%$) or not practically equivalent and inconclusive ($r\delta_M \geq 30\%$). While all the cited publications in the table correctly stated that no difference was observed between the control and experiment group, most results were still used indirectly as evidence of negligible effect size (see NE column) either in the text or as a secondary negative control. With these thresholds, scientists could not use results with $r\delta_M \geq 30\%$ as evidence of absence of effect. Instead, they could choose to present the data with an ambiguous interpretation or collect additional samples in an attempt to clarify the interpretation for $r\delta_M$.

As a second example, a similar case study examines the practical equivalence of therapeutic interventions independent of reducing plaque size (Fig. 5A-B, Table S5). Similar to measuring total cholesterol, plaque size is measured across units that span orders of magnitude. A review of positive results of plaque size reduction (Table S6) could yield a threshold of 40% for a negligible effect size. Using that threshold could separate results that are practically equivalent ($r\delta_M < 40\%$) or not practically equivalent and inconclusive ($r\delta_M \geq 40\%$). Inconclusive results could not be interpreted as evidence of negligible effect size, as the authors intended in most cases.

**Discussion**

We have proposed two statistics, $\delta_M$ and $r\delta_M$, that assess the evidence of negligible effect size by quantifying null strength. Both statistics support hypothesis testing and were the only candidate statistics that exhibited lower than random error in comparing null strength across all investigations. We demonstrated with applied examples how researchers can use $r\delta_M$ to assess practical equivalence by evaluating both the negligibility and null strength of experiment results. This statistic summarizes results as a simple percentage with a simple interpretation (largest percent change between mean of control group and experiment group supported by the data). We illustrate that researchers can apply a threshold and identify practically equivalent results. Critically, researchers can apply different thresholds based on their differing opinions or research applications without having to re-compute $r\delta_M$: the value of the statistic remains unchanged if different thresholds are used in Fig. 4 and 5. Results that are designated as practically equivalent can be used to falsify scientific hypotheses or offer contrary evidence to related positive results.

Our hypothesis testing approach in Fig. 1 of examining the overlap between an interval and null region aligns closely with the procedure used to calculate second generation p-values (*14*, *15*). Indeed, both procedures would designate the same results as practically equivalent (i.e., full support of the null hypothesis) if the same threshold and intervals were used. The advantage of using $\delta_M$ is that null strength between results can be compared regardless of their designation for practical equivalence, and different null regions can be used for hypothesis testing after the data is reported. In contrast to our statistic, a collection of practically equivalent results would all have a second-generation p-value of 1, so their null strength cannot be compared. The second-generation p-value would also require recalculation if a different null region is specified. There are similar limitations with using the Bayes Factor and two one-sided t-test p-values. We note that the treatment of these statistics in Figures 2-5 was generous because the null interval regions were fixed across simulated experiments. In practice the extent of these null regions would vary, and these candidate statistics would not be comparable across studies.

Alternatively, some fields may report the confidence or credible interval of the population DM to aid in interpreting the practical equivalence of results. Reporting these intervals is helpful because they examine both statistical and practical significance by estimating a range for the effect size specified by the credible or confidence level. Yet comparing the evidence of practical equivalence with intervals must consider the width and location of the interval when there is no clear method to combine them. We solve this issue by developing a statistic that collapses an interval into a single value.

For experimental results, the presence or lack of negligible effect size should be interpreted in the context of related results. We recommend that $r\delta_M$ should be the default statistic used to evaluate the strength of practically equivalent results since it allows for comparisons between a broader range of related experiments than $\delta_M$. However, $\delta_M$ would be more appropriate for cases where the mean of the control group is close to zero and reporting the percent difference in means is spurious (e.g., see Fig 5A, last study, where $r\delta_M$ of 450% is more than twice the value of $\delta_M$ divided by the control mean: $\delta_M/\overline{x}=180\%$), or for comparisons between experiments where the control group mean is not expected to change. Reporting the $\delta_M$ or $r\delta_M$ encourages high quality results because it rewards the use of larger sample sizes, higher quality measurement techniques, and more rigorous experiment design. Additionally, assessing null results with the $r\delta_M$ may reduce publication bias against null results (*18*), mitigate the File Drawer problem by encouraging their publication, and increase scientific rigor by allowing results with high null strength to serve as strong evidence of negligible effect size. If combined with another statistic that could quantify practical significance in the same manner, $r\delta_M$ would provide the foundation for a complete method of data analysis that could be used in place of p-values.

**Limitations of the Study**

This statistic requires the assumption of normality. Other versions of this statistic must be developed to analyze cases where this assumption is false. Using this statistic will change how scientists report results. Further research should establish specific conventions for the wording when reporting this statistic to minimize miscommunication between scientists.

**Author Contributions:**

Investigation, Writing- Original draft, Visualization, Data Curation: BAC.

Software: BAC, TRB.

Methodology, Formal Analysis: BAC, TRB, TZ.

Conceptualization: KAJ, BAC.

Writing- Reviewing and Editing: BAC, TRB, TZ, KAJ, HS, PEB.

Supervision: PEB.

**Acknowledgements:** we would like to thank the Biocomplexity Institute at UVA for their invaluable feedback.  This work was funded by PEB's endowment for the School of Data Science, University of Virginia.

**Declaration of Interests:**  The authors declare no competing interests.

**Data and Material Availability**: code used to generate all data and figures is written in R and available at: https://github.com/bac7wj/ACES.

**Figures and Legends**

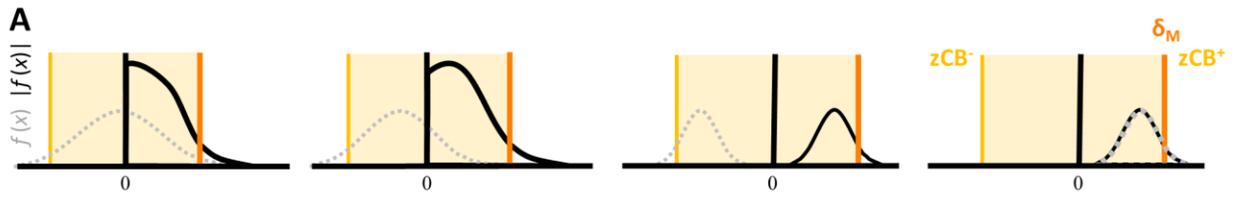

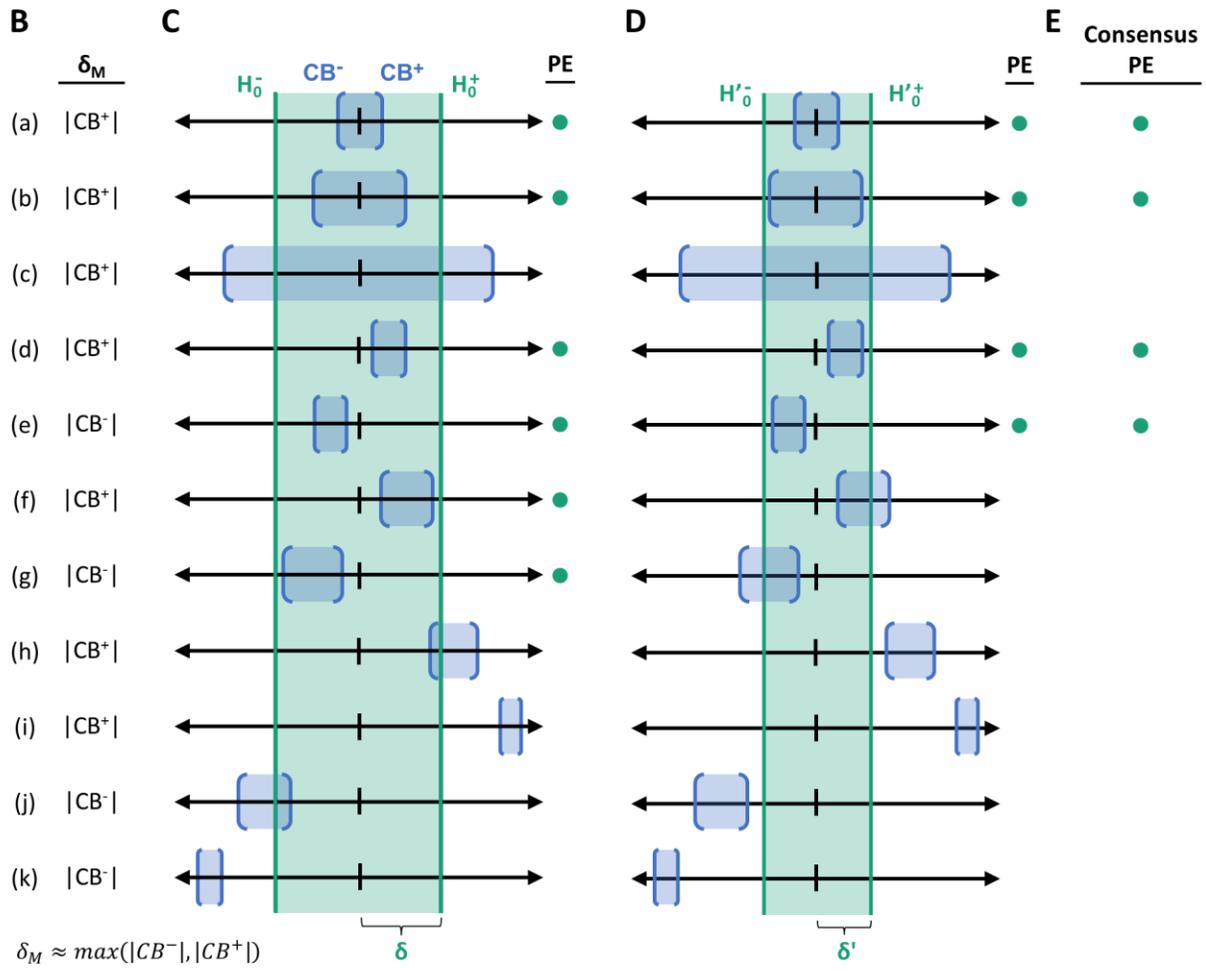

*Fig. 1*: **Using the most difference in means to reach consensus for practical equivalence.** (**A**) Several illustrations of the most difference in means statistic (orange line, $\delta_M$) as the upper bound of a zero-centered credible interval (yellow fill, $zCB^{\pm}$) with posterior of difference in means (dashed grey) and absolute difference in means (solid black). (**B**) A series of hypothetical two-sample experiments with the value of $\delta_M$ approximated as the maximum of the absolute lower and upper credible bounds of the difference in means (blue, $CB^{\pm}$). (**C**) A scientist specifies a threshold ($\delta$, enclosing green null interval $H_0^{\pm}$) to designate each result as practically equivalent (PE, CB interval fully within null interval, $\delta_M < \delta$, green dot), or not practically equivalent (CB interval partially or completely outside of null interval, $\delta_M \geq \delta$). (**D**) Repeated analysis can be done by a second scientist with a different threshold ($\delta$') with no recalculation of $\delta_M$ required. (**E**) Consensus is reached when both scientists designate a result as practically equivalent (green dot).

## Raw Null Strength Measures

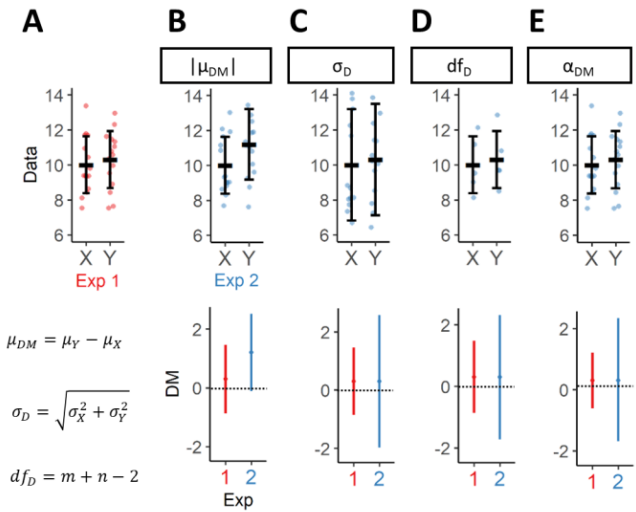

$$\mu_{DM} = \mu_Y - \mu_X$$

$$\sigma_D = \sqrt{\sigma_X^2 + \sigma_Y^2}$$

$$df_D = m + n - 2$$

## Relative Null Strength Measures

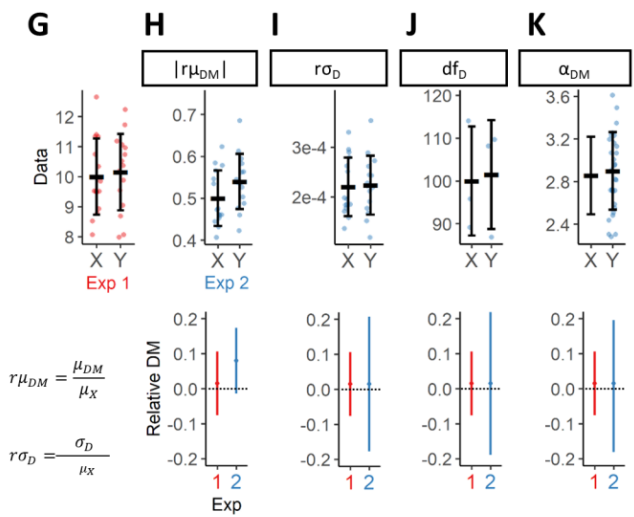

$$r\mu_{DM} = \frac{\mu_{DM}}{\mu_X}$$

$$r\sigma_D = \frac{\sigma_D}{\mu_X}$$

### F — Raw Null Strength Measures

| Statistic | $|\mu_{DM}|$ | $r\sigma_D$ | $df_D$ | $\alpha_{DM}$ |
|---|---|---|---|---|
| $\bar{x}_{DM}$ | -1.00 | -0.74 | -0.24 | -0.41 |
| $r\bar{x}_{DM}$ | +0.28 | +0.20 | +0.06 | -0.42 |
| $s_{DM}$ | -0.45 | -1.00 | -1.00 | +0.51 |
| $rs_{DM}$ | -0.45 | -0.43 | -1.00 | +0.48 |
| Bf | +1.00 | +0.14 | -0.79 | -0.42 |
| $p_N$ | +0.95 | +0.08 | -0.03 | +0.46 |
| * $p_E$ | -0.39 | -1.00 | -0.99 | -0.89 |
| $p_\delta$ | +0.39 | +0.72 | +0.81 | -0.99 |
| CD | -1.00 | -0.16 | -0.28 | -0.73 |
| * $\delta_M$ | -1.00 | -1.00 | -1.00 | -1.00 |
| $r\delta_M$ | +0.78 | -0.33 | | -0.99 |
| Rnd | +0.27 | +0.17 | -0.07 | -0.32 |

### L — Relative Null Strength Measures

| Statistic | $|r\mu_{DM}|$ | $r\sigma_D$ | $df_D$ | $\alpha_{DM}$ |
|---|---|---|---|---|
| $\bar{x}_{DM}$ | -0.04 | +0.02 | +0.12 | +0.52 |
| $r\bar{x}_{DM}$ | -1.00 | -0.60 | +0.18 | +0.46 |
| $s_{DM}$ | +1.00 | -0.47 | -1.00 | -0.10 |
| $rs_{DM}$ | +0.91 | -1.00 | -1.00 | -0.11 |
| Bf | -0.73 | +0.44 | -0.38 | +0.11 |
| $p_N$ | +1.00 | +0.08 | -0.25 | +0.07 |
| * $p_E$ | -1.00 | -0.98 | -0.98 | -0.86 |
| $p_\delta$ | +1.00 | -0.35 | +0.68 | +0.00 |
| CD | -1.00 | -0.08 | +0.09 | +0.55 |
| $\delta_M$ | +1.00 | -0.43 | -1.00 | -0.98 |
| * $r\delta_M$ | -1.00 | -1.00 | -1.00 | -0.98 |
| Rnd | +0.14 | +0.15 | +0.21 | -0.56 |

Spearman $\rho$ of Statistic Vs. Measure
Towards Higher Null Strength

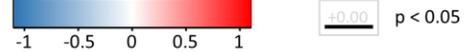

-1   -0.5   0   0.5   1

$p < 0.05$

*Fig. 2*: **Covariation of candidate statistics with measures of null strength.** (**A**) Data from a simulated experiment (red, Exp 1) with a control group (X) and experiment group (Y) acting as a reference to illustrate the measures of null strength. (**B-E**) Simulated experiment data (Exp 2, blue) with lower null strength of difference in means (DM) than Exp 1 (lower panel) via (**B**) increased difference in means, (**C**) increased standard deviation of the difference, (**D**) decreased degrees of freedom, and (**E**) decreased credible level (upper: error bars are standard deviation, lower: error lines are 95% credible interval of the difference in means). (**F**) Heatmap of Spearman ρ of candidate statistics' mean versus each raw null strength measure altered towards higher null strength across population configurations. (**G**) Simulated data from an experiment (red, Exp 1) acting as a reference to illustrate the measures of relative null strength. (**H-K**) Simulated experiment data (Exp 2, blue) with lower relative null strength of relative difference in means than Exp 1 (lower panel) via (**H**) increased relative difference in means, (**I**) increased relative standard deviation of the difference, (**J**) decreased degrees of freedom, and (**K**) decreased significance level. (**L**) Heatmap of Spearman ρ of candidate statistics' mean versus each relative null strength measure altered towards higher null strength across population configurations. Asterisk denotes candidate statistic with all correlations significant and in same direction, underline denotes $p < 0.05$ for bootstrapped Spearman correlation, color displayed for significant correlations only). Abbreviations: $\bar{x}_{DM}$, $s_{DM}$, $r\bar{x}_{DM}$, $rs_{DM}$: mean, standard deviation, relative mean, and relative standard deviation of difference in sample means. CD: Cohen's d; $P_N$: null hypothesis testing p-value; $P_E$: TOST equivalence p-value; $P_\delta$: second generation p-value; BF: Bayes Factor; Rnd: random 50/50 guess.

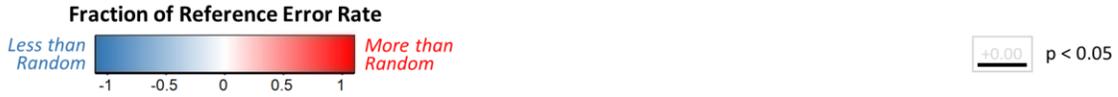

**Fraction of Reference Error Rate**

*Less than Random* — *More than Random*
-1  -0.5  0  0.5  1

| +0.00 | p < 0.05

## A

### Comparison Error Rate for Raw Null Strength

| | Null Results: $|t_{statistic} / t_{critical}| \leq 1$ | | | | | | | | Positive Results: $|t_{statistic} / t_{critical}| > 1$ | | | | | | | |
| --- | --- | --- | --- | --- | --- | --- | --- | --- | --- | --- | --- | --- | --- | --- | --- | --- |
| | Individual | | | | Simultaneous | | | | Individual | | | | Simultaneous | | | |
| | $|\mu_{DM}|$ | $r\sigma_D$ | $df_D$ | $r\alpha_{DM}$ | $|\mu_{DM}|$ | $r\sigma_D$ | $df_D$ | $\alpha_{DM}$ | $|\mu_{DM}|$ | $r\sigma_D$ | $df_D$ | $\alpha_{DM}$ | $|\mu_{DM}|$ | $r\sigma_D$ | $df_D$ | $\alpha_{DM}$ |
| $\bar{x}_{DM}$ | -1.00 | -0.15 | -0.14 | +0.01 | -1.00 | -0.12 | -0.17 | +0.07 | -1.00 | -0.00 | -0.00 | +0.02 | -1.00 | -0.02 | -0.01 | +0.06 |
| $r\bar{x}_{DM}$ | -0.09 | -0.01 | -0.02 | -0.01 | -0.10 | +0.06 | -0.25 | +0.05 | -0.00 | -0.00 | +0.01 | -0.02 | -0.03 | +0.05 | -0.12 | +0.07 |
| $s_{DM}$ | +0.04 | -1.00 | -1.00 | -0.00 | -0.05 | -1.00 | -0.98 | +0.01 | +0.01 | -1.00 | -1.00 | -0.01 | -0.01 | -1.00 | -1.00 | -0.03 |
| $rs_{DM}$ | +0.13 | +0.02 | +0.01 | -0.01 | -0.02 | +0.08 | -0.25 | +0.04 | +0.02 | -0.01 | +0.01 | -0.02 | -0.03 | +0.05 | -0.12 | +0.07 |
| Bf | -0.99 | -0.15 | -0.33 | +0.01 | -0.84 | +0.20 | +0.16 | +0.05 | -0.98 | +0.66 | +0.94 | +0.01 | +0.04 | -0.01 | +1.50 | +0.04 |
| $p_N$ | -1.00 | +0.09 | +0.11 | +0.01 | -0.93 | +0.23 | +0.24 | +0.05 | -0.98 | +0.68 | +0.88 | +0.01 | -0.14 | +0.40 | +1.30 | +0.04 |
| $p_E$ | -1.00 | -0.25 | -0.47 | +0.01 | -0.17 | +0.06 | -0.16 | -0.86 | -0.99 | -0.36 | +0.54 | -1.00 | -0.05 | +0.08 | -0.10 | +0.01 |
| $p_\delta$ | -0.60 | -0.43 | -0.51 | -0.19 | -0.37 | +0.19 | -0.03 | +0.39 | -0.60 | -0.35 | +0.38 | +0.10 | -0.23 | +0.67 | +0.52 | +0.71 |
| CD | -1.00 | +0.09 | -0.14 | +0.01 | -0.94 | +0.22 | -0.20 | +0.07 | -0.98 | +0.67 | -0.03 | +0.01 | -0.41 | +1.89 | -0.11 | -0.02 |
| * $\delta_M$ | -0.98 | -0.64 | -0.68 | -1.00 | -0.51 | -0.62 | -1.00 | -1.00 | -1.00 | -0.39 | -0.52 | -0.97 | -0.56 | -0.65 | -0.82 | -1.00 |
| $r\delta_M$ | +0.13 | +0.02 | +0.01 | -0.01 | -0.02 | +0.08 | -0.25 | +0.04 | +0.02 | -0.01 | +0.01 | -0.02 | -0.03 | +0.05 | -0.12 | +0.07 |
| Rnd | -0.00 | -0.00 | -0.00 | -0.01 | +0.02 | -0.02 | +0.01 | +0.01 | -0.00 | -0.01 | +0.00 | -0.01 | -0.01 | +0.00 | +0.00 | -0.00 |

## B

### Comparison Error Rate for Relative Null Strength

| | Null Results: $|t_{statistic} / t_{critical}| \leq 1$ | | | | | | | | Positive Results: $|t_{statistic} / t_{critical}| > 1$ | | | | | | | |
| --- | --- | --- | --- | --- | --- | --- | --- | --- | --- | --- | --- | --- | --- | --- | --- | --- |
| | Individual | | | | Simultaneous | | | | Individual | | | | Simultaneous | | | |
| | $|r\mu_{DM}|$ | $r\sigma_D$ | $df_D$ | $r\alpha_{DM}$ | $|r\mu_{DM}|$ | $r\sigma_D$ | $df_D$ | $\alpha_{DM}$ | $|r\mu_{DM}|$ | $r\sigma_D$ | $df_D$ | $\alpha_{DM}$ | $|r\mu_{DM}|$ | $r\sigma_D$ | $df_D$ | $\alpha_{DM}$ |
| $\bar{x}_{DM}$ | -0.18 | -0.03 | -0.09 | +0.01 | -0.42 | -0.01 | -0.00 | +0.02 | -0.01 | +0.03 | -0.00 | -0.01 | -0.04 | +0.01 | -0.03 | +0.03 |
| $r\bar{x}_{DM}$ | -1.00 | -0.17 | -0.23 | -0.00 | -0.97 | -0.10 | -0.14 | +0.02 | -1.00 | -0.00 | +0.00 | -0.01 | -1.00 | +0.03 | -0.04 | +0.17 |
| $s_{DM}$ | +0.01 | -0.02 | -0.01 | +0.01 | -0.08 | +0.01 | +0.01 | +0.03 | +0.02 | +0.00 | -0.02 | +0.00 | -0.03 | -0.02 | +0.00 | +0.04 |
| $rs_{DM}$ | -0.01 | -1.00 | -0.94 | +0.01 | -0.08 | -1.00 | -1.00 | -0.01 | -0.06 | -1.00 | -1.00 | +0.00 | -0.06 | -1.00 | -1.00 | -0.09 |
| Bf | -1.00 | +0.38 | -0.32 | -0.00 | -1.00 | +0.27 | -0.01 | +0.03 | -0.96 | +0.53 | +1.52 | -0.00 | -0.06 | +0.32 | +1.59 | +0.06 |
| $p_N$ | -1.00 | +0.43 | +0.18 | +0.01 | -0.90 | +0.37 | +0.33 | +0.02 | -0.96 | +0.55 | +1.28 | +0.01 | -0.21 | +0.57 | +1.37 | +0.07 |
| $p_E$ | -0.99 | +0.43 | -1.00 | -1.00 | -0.11 | -0.11 | +0.04 | -1.00 | -0.05 | +0.21 | +1.00 | -0.36 | +0.02 | +0.01 | -0.01 | -0.05 |
| $p_\delta$ | -0.99 | +0.16 | -0.03 | -0.42 | -0.56 | +0.28 | +0.28 | +0.31 | -0.02 | +0.20 | +0.35 | +0.33 | -0.14 | +0.38 | +0.46 | +0.33 |
| CD | -1.00 | +0.43 | -0.24 | +0.00 | -0.95 | +0.35 | -0.16 | -0.00 | -0.96 | +0.55 | -0.06 | -0.00 | -0.36 | +1.61 | -0.07 | +0.25 |
| $\delta_M$ | +0.01 | -0.02 | -0.01 | +0.01 | -0.08 | +0.01 | +0.01 | +0.02 | +0.02 | +0.00 | -0.02 | +0.00 | -0.04 | +0.01 | +0.03 | +0.03 |
| * $r\delta_M$ | -0.97 | -0.99 | -0.83 | -0.66 | -0.47 | -0.75 | -0.87 | -0.56 | -0.99 | -0.33 | -0.58 | -1.00 | -0.65 | -0.61 | -0.78 | -1.00 |
| Rnd | +0.00 | +0.01 | +0.00 | -0.00 | +0.01 | +0.01 | -0.00 | +0.00 | -0.00 | -0.01 | +0.00 | -0.00 | -0.01 | +0.01 | +0.01 | -0.01 |

*Fig. 3*: **Comparison error rates of candidate statistics in identifying higher null strength between results.** (**A**) Heatmap of comparison error rates for each candidate statistic across raw null strength measures for identifying which of two results have higher raw null strength. (**B**) Heatmap of comparison error rates for each candidate statistic across relative null strength measures for identifying which of two results have higher relative null strength. Blue fill denotes comparison error less than random, red denotes greater than random, and white is no different than random. Numerical label in cells are comparison error rates from random behavior scaled to the lowest error rate for each column. Underlined numbers denote a comparison error rate that is statistically different than random. Investigations alter one measure of null strength as independent variable (Individual) or several at once (Simultaneous) to serve as ground truth. Investigations are separated between population configurations associated with null results (expected t-ratio $\leq 1$) and critical results (expected t-ratio $> 1$). See Fig. 2 for abbreviations.

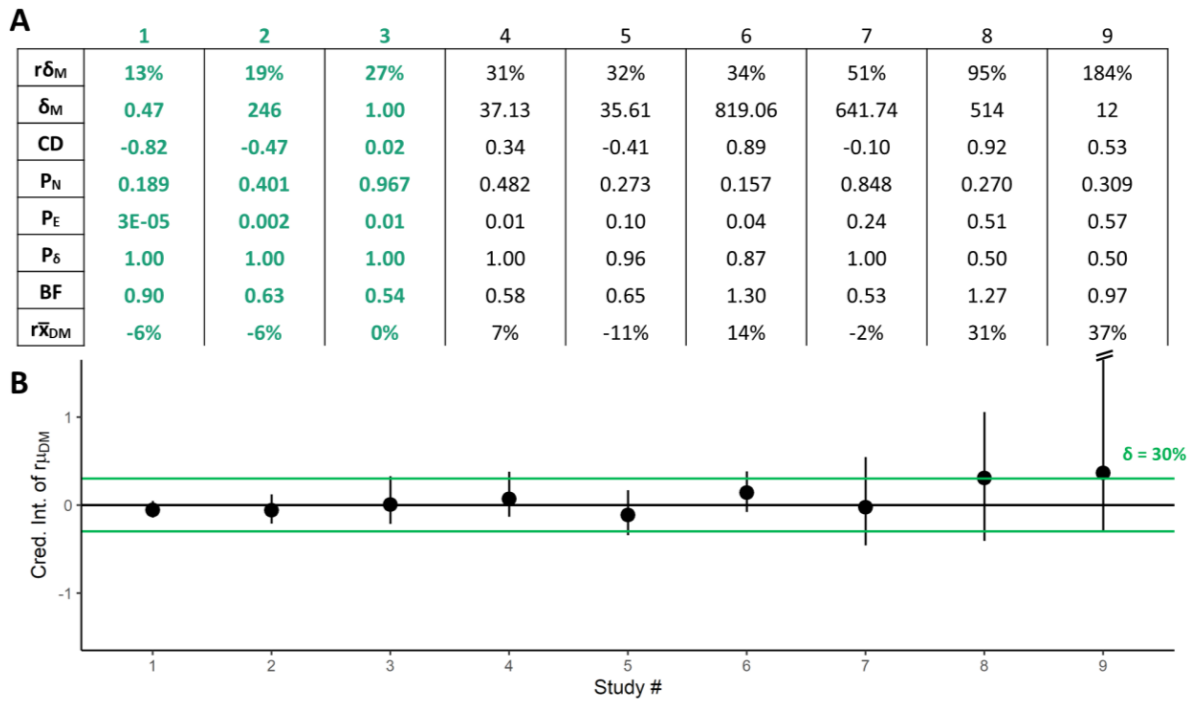

| | 1 | 2 | 3 | 4 | 5 | 6 | 7 | 8 | 9 |
|---|---|---|---|---|---|---|---|---|---|
| $r\delta_M$ | 13% | 19% | 27% | 31% | 32% | 34% | 51% | 95% | 184% |
| $\delta_M$ | 0.47 | 246 | 1.00 | 37.13 | 35.61 | 819.06 | 641.74 | 514 | 12 |
| CD | -0.82 | -0.47 | 0.02 | 0.34 | -0.41 | 0.89 | -0.10 | 0.92 | 0.53 |
| $P_N$ | 0.189 | 0.401 | 0.967 | 0.482 | 0.273 | 0.157 | 0.848 | 0.270 | 0.309 |
| $P_E$ | 3E-05 | 0.002 | 0.01 | 0.01 | 0.10 | 0.04 | 0.24 | 0.51 | 0.57 |
| $P_\delta$ | 1.00 | 1.00 | 1.00 | 1.00 | 0.96 | 0.87 | 1.00 | 0.50 | 0.50 |
| BF | 0.90 | 0.63 | 0.54 | 0.58 | 0.65 | 1.30 | 0.53 | 1.27 | 0.97 |
| $r\bar{x}_{DM}$ | -6% | -6% | 0% | 7% | -11% | 14% | -2% | 31% | 37% |

***Fig. 4***: **Interpreting null results of total plasma cholesterol in atherosclerosis research.** (**A**) Table of candidate statistics summarizing null results from a collection of studies (columns 1-9, practically equivalent results according to $r\delta_M$ highlighted in green). (**B**) For visual reference, 95% credible interval of the relative difference in means (estimated with Monte-Carlo sampling of posterior of $r\mu_{DM}$ with a noninformative uniform prior). Null region interval ($\delta$) was set to [-30%, +30%] of control sample mean for $r\delta_M$, $P_E$, $P_\delta$, and BF. Credible intervals are Bonferroni adjusted according to each study design (see Table S3 for details and citation for each study). See Fig. 2 for candidate statistic abbreviations.

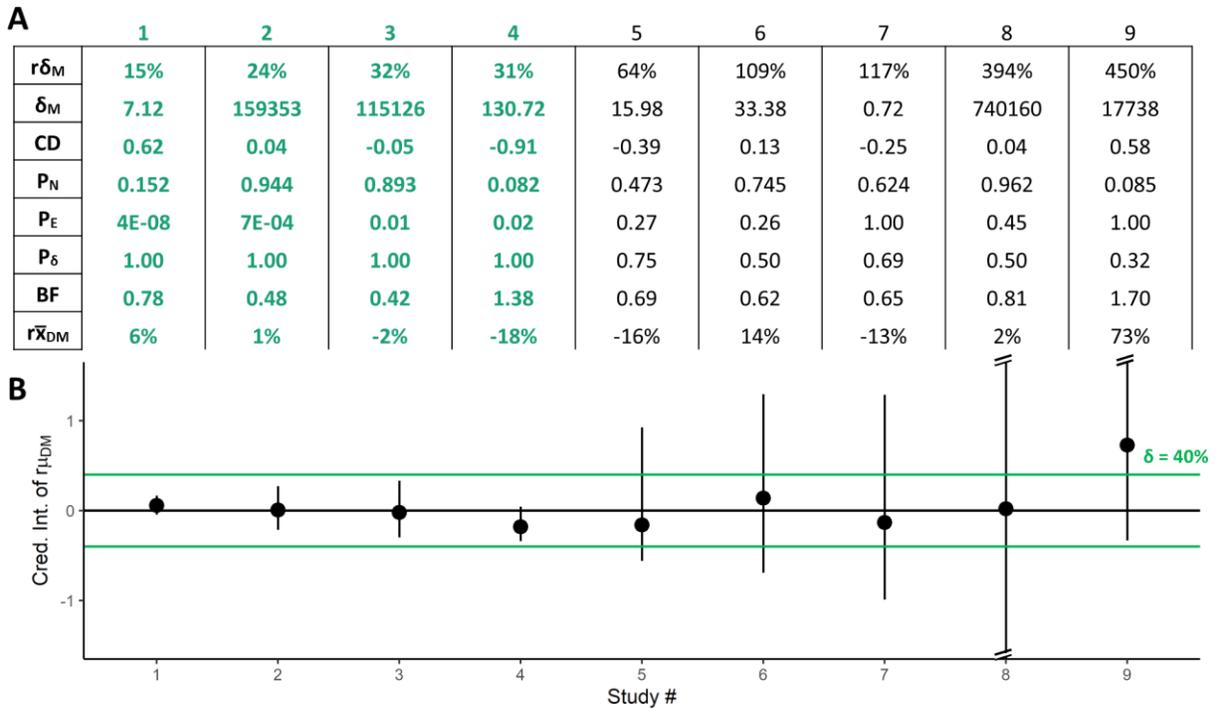

**A**

| | 1 | 2 | 3 | 4 | 5 | 6 | 7 | 8 | 9 |
|---|---|---|---|---|---|---|---|---|---|
| $r\delta_M$ | 15% | 24% | 32% | 31% | 64% | 109% | 117% | 394% | 450% |
| $\delta_M$ | 7.12 | 159353 | 115126 | 130.72 | 15.98 | 33.38 | 0.72 | 740160 | 17738 |
| CD | 0.62 | 0.04 | -0.05 | -0.91 | -0.39 | 0.13 | -0.25 | 0.04 | 0.58 |
| $P_N$ | 0.152 | 0.944 | 0.893 | 0.082 | 0.473 | 0.745 | 0.624 | 0.962 | 0.085 |
| $P_E$ | 4E-08 | 7E-04 | 0.01 | 0.02 | 0.27 | 0.26 | 1.00 | 0.45 | 1.00 |
| $P_\delta$ | 1.00 | 1.00 | 1.00 | 1.00 | 0.75 | 0.50 | 0.69 | 0.50 | 0.32 |
| BF | 0.78 | 0.48 | 0.42 | 1.38 | 0.69 | 0.62 | 0.65 | 0.81 | 1.70 |
| $r\bar{x}_{DM}$ | 6% | 1% | -2% | -18% | -16% | 14% | -13% | 2% | 73% |

*Fig. 5*: **Interpreting null results of arterial plaque size in atherosclerosis research.** (**A**) Table of candidate statistics summarizing null results from a collection of studies (columns 1-9, practically equivalent results according to $r\delta_M$ highlighted in green). (**B**) For visual reference, 95% credible interval of the relative difference in means (estimated with Monte-Carlo sampling of posterior for $r\mu_{DM}$ with a noninformative uniform prior). Null region interval ($\delta$) was set to [-40%, +40%] of control sample mean for $r\delta_M$, $P_E$, $P_\delta$, and BF. Credible intervals are Bonferroni adjusted according to each study design (see Table S5 for details and citation for each study). See Fig. 2 for candidate statistic abbreviations.

**Tables**

*Table 1: Measures of Null strength*

| Measure | Scale | Equation | Higher Null Strength |
|---|---|---|---|
| $\|\mu_{DM}\|$ | Raw | $\|\mu_{DM}\| = \|\mu_Y - \mu_X\|,$ | − |
| $\sigma_D$ | Raw | $\sigma_D = \sqrt{\sigma_X^2 + \sigma_Y^2}$ | − |
| $df_D$ | Raw, Relative | $df_D = m + n - 2$ | + |
| $\alpha_{DM}$ | Raw, Relative | | + |
| $\|r\mu_{DM}\|$ | Relative | $\|r\mu_{DM}\| = \left\|\dfrac{\mu_{DM}}{\mu_X}\right\|$ | − |
| $r\sigma_D$ | Relative | $r\sigma_D = \dfrac{\sigma_D}{\mu_X}$ | − |

*Note: Higher null strength column indicates the direction of change for each measure to increase null strength when other measures held constant. Abbreviations: D, difference distribution of X and Y.*

**Materials and Methods**

*Explanation of Raw Measures of Null Strength*

We defined raw null strength as an estimate of how large $\|\mu_{DM}\|$ could be based on sample data. From a Bayesian perspective, this can be represented with an upper quantile of a posterior distribution summarizing $\|\mu_{DM}\|$. Therefore, we must consider not only the location, but also the dispersion of the distribution summarizing $\|\mu_{DM}\|$ because both can change its upper quantiles.

Based on our definition of raw null strength, higher null strength is found with lower values of $|\mu_{DM}|$ with all other measures held constant, illustrated with higher null strength from experiment 1 with its credible interval for $\mu_{DM}$ centered closer to zero than experiment 2 (Fig 2A, B). Higher null strength is also found with lower values of $\sigma_{DM}$ with all other measures held constant since it suggests a smaller upper bound for $|\mu_{DM}|$. Since $\sigma_{DM}$ is influenced by both the standard deviations and sample sizes of both groups, the contributions of each can be independently characterized with the standard deviation ($\sigma_D$) and degrees of freedom ($df_D$) of the difference between observations from group X and Y (i.e., D = Y - X).

$$\sigma_D = \sqrt{\sigma_X^2 + \sigma_Y^2} \qquad (16)$$

$$df_D = m + n - 2 \qquad (17)$$

There is higher null strength with lower values of $\sigma_D$ (contributing to $\sigma_{DM}$ in the numerator) with all other measures held constant, illustrated with higher null strength from experiment 1 with its narrower credible interval (Fig. 2C). There is also higher null strength with higher values of $df_D$ (contributing to $\sigma_{DM}$ in the denominator) with all other measures held constant, illustrated with higher null strength from experiment 1 with its narrower credible interval (Fig. 2D). In addition to $\sigma_{DM}$ indicating how large the range of $\mu_{DM}$ could be, the specified credible level ($\alpha_{DM}$) also effects the uncertainty associated with the comparison (often adjusted for experiments with multiple comparisons). There is higher null strength with lower values of $\alpha_{DM}$ with all other measures held constant because there is an increase in the range of possible values for $\mu_{DM}$, illustrated with higher null strength from experiment 1 with its narrower credible interval (Fig. 2E).

We have identified $|\mu_{DM}|$, $\sigma_D$, $df_D$, and $\alpha_{DM}$ as measures of null strength (Table 1) by illustrating how changes to each of these measures in isolation leads to known changes to null strength. Since the value of these measures can be altered independently, each of these measures can be altered as an independent measure to test the effectiveness of candidate statistics in quantifying null strength. An effective statistic should be able to identify results with higher null strength across all four of these measures.

### *Explanation of Relative Measures of Null strength*

To quantify relative null strength, we extend the measures of null strength into units relative to the mean of the control sample. The relative difference in means ($r\mu_{DM}$) and relative standard deviation ($r\sigma_{DM}$) are normalized by the mean of the control group:

$$r\mu_{DM} = \frac{\mu_{DM}}{\mu_X} \qquad\qquad (18)$$

$$r\sigma_{DM} = \frac{\sigma_{DM}}{\mu_X}. \qquad\qquad (19)$$

We quantify relative null strength by estimating the upper bound of the magnitude of $r\mu_{DM}$, where smaller values exhibit higher null strength.

Lower relative null strength is found with lower values of the magnitude of $r\mu_{DM}$ (abbreviated as $|r\mu_{DM}|$) with all other measures held constant, illustrated with experiment 1 having a credible interval for $r\mu_{DM}$ centered closer to zero (Fig 2G, H). Lower relative null strength is also found with lower values $r\sigma_{DM}$ with all other measures held constant. Since $r\sigma_{DM}$ is influenced by both the relative standard deviations and sample sizes of both groups, the contributions of each can be independently characterized with the relative standard deviation ($r\sigma_D$) and degrees of freedom of the difference between observations:

$$r\sigma_D = \frac{\sigma_D}{\mu_X}. \qquad\qquad (20)$$

There is lower relative null strength with lower values of $r\sigma_D$ with all other measures held constant, illustrated with higher null strength from experiment 1 with its narrower credible interval (Fig. 2I). There is lower relative null strength with higher values of $df_D$ (contributing to $\sigma_{DM}$ in the denominator) with all other measures held constant, illustrated with higher null strength from experiment 1 with its narrower credible interval (Fig. 2J). In addition to $r\sigma_{DM}$ indicating how large the range of $r\mu_{DM}$ could be, the specified credible level ($\alpha_{DM}$) also effects the uncertainty associated with the comparison (often adjusted for experiments with multiple comparisons). There is higher null strength with lower values of $\alpha_{DM}$ with all other measures held constant because there is an increase in the range of possible values for $r\mu_{DM}$, illustrated with higher null strength from experiment 1 with its narrower credible interval (Fig 2K).

We have identified $|r\mu_{DM}|$, $r\sigma_D$, $df_D$, and $\alpha_{DM}$ as measures of relative null strength (Table 1) by illustrating changes to each of these measures in isolation leads to known changes to relative null strength. Since the value of these measures can be altered independently, each of these measures can be varied as independent variables to test the effectiveness of candidate statistics in quantifying relative null strength. An effective statistic should be able to identify results with lower relative null strength across all four of these measures.

***Integrated Risk Assessment of Null Strength***

Identifying results with higher null strength is a critical feature for assessing practical equivalence. Our risk assessment is designed to benchmark the efficacy of various candidate statistics in determining which of two experiments has higher null strength and deemed more noteworthy. In many cases, such a determination is difficult since null strength is a function of several parameters (see Table 1). To simulate instances where it is clear which experiment has higher null strength, we hold all population parameters constant except those that alter a specified null strength measure (referred to as the independent measure). Using this strategy, we can then benchmark performance of the candidate statistics in determining higher null strength for each measure of null strength in isolation. Since the null strength measures represent known instances where null strength changes, we set the criterion that a successful statistic must predict higher null strength at a rate better than random for every measure of null strength. We test for this criterion with simulations that use both frequentist and Bayesian approaches to assess risk (in this case, risk is defined as the probability of incorrectly predicting which of two results has higher null strength).

Given the nomenclature defined in the background section (Bayesian Summary of Difference in Means), let $\theta$ be a population parameter configuration for a hypothetical experiment 1. Specifically, $\theta$ is a vector of population parameters and the credible level specified for experiment 1 required to simulate sample data from a control group X and experiment group Y.

$$\theta = (\mu_X, \mu_Y, \sigma_X^2, \sigma_Y^2, \alpha_{DM}),$$ *(21)*

And let $\theta'$ be a population configuration from a second simulated experiment. For a pair of population configurations $\theta$ and $\theta'$, we determine which is more noteworthy based on the ground truth determined by the independent measure of null strength. We then draw samples from these population configurations and use various candidate statistics to predict which experiment is most noteworthy and compare this prediction against the ground truth.

To establish ground truth, we defined a loss function for each null strength measure that designates whether experiment 1 or experiment 2 had higher null strength (Table S2). These loss functions assume that all other measures of null strength are held constant between the two experiments. The loss functions compare the ground truth from the null strength measure ("ground truth designation") versus the prediction from a candidate statistic ("predicted designation").

A value of 1 from the ground truth and prediction designations denote experiment 1 having higher null strength than experiment 2. A value of 0 from the loss function denotes that the candidate statistic agrees with the ground truth (inequalities for candidate designations are switched in loss functions).

Using these loss functions, we can approximate, using the Monte Carlo method, the frequentist risk (*16*) for a single population configuration by calculating the mean loss over *M* samples.

$$\widehat{\mathbb{P}}_{\theta,\theta'}(|\delta(x,y,\alpha_{DM})| < |\delta(x^{'},y^{'},\alpha^{'}_{DM})|) \coloneqq \frac{1}{M}\sum_{i=1}^{M}Loss\big(x_i,y_i,x^{'}_i,y^{'}_i,\theta,\theta'\big). \tag{22}$$

We define this evaluation of frequentist risk as comparison error, interpreted as the probability that a candidate statistic will incorrectly assign greater noteworthiness when comparing two experiments.

Since frequentist risk is a function of the population configuration, analyzing risk at one configuration will not be representative of general performance. To explore general trends across the parameter space, we averaged comparison errors from many different parameter configurations.

$$\frac{1}{N}\sum_{i=1}^{N}\widehat{\mathbb{P}}_{\theta_i,\theta'_i}(|\delta(x,y,\alpha_{DM})| < |\delta(x^{'},y^{'},\alpha^{'}_{DM})|) \approx \mathbb{E}_{\theta_i,\theta'_i}[\widehat{\mathbb{P}}_{\theta_i,\theta'_i}|\delta(x,y,\alpha_{DM})| < |\delta(x^{'},y^{'},\alpha^{'}_{DM})|] \tag{23}$$

This strategy of averaging frequentist risks follows the process for assessing integrated Bayesian risk (*16*). However, in an integrated Bayesian risk assessment, the population configurations must be generated randomly from a specific prior. We do not follow this strategy because our investigation requires a more structured characterization of trends within the parameter space.

A Bayesian risk assessment would simply generate a single value that globally summarizes risk, where a lower risk is considered better. However, our chosen criterion is that a successful statistic must successfully predict noteworthiness from changes to every null strength measure in isolation. To accomplish this, the risk assessment must individually investigate each measure of null strength so a direct relationship with candidate statistics' performance can be ascertained. Generating population configurations from a specific prior could not achieve these objectives. Instead, we must carefully generate curated population configurations (See Supplementary Materials and Methods for more details).

**Supplementary Materials**

Materials and Methods

Supplementary Text

Figs. S1 to S14

Tables S1 to S6

# Supplementary Materials for

## The Most Difference in Means: A Statistic for the Strength of Null and Near-Zero Results

Bruce A. Corliss, Taylor R. Brown, Tingting Zhang, Kevin A. Janes, Heman Shakeri, Philip E. Bourne

Correspondence to: bac7wj@virginia.edu

**This PDF file includes:**

Supplementary Materials and Methods
Figs. S1 to S14
Tables S1 to S6
Supplementary References

**Supplementary Materials and Methods**

*Literature Search*

The tables summarizing results in Fig. 4 and 5 were compiled based on a literature search using Pubmed, Google Scholar, and Google search. Included results were limited to papers that were indexed on Pubmed. Papers were identified based on searches with combinations of the following keywords:

Total cholesterol example: atherosclerosis, total cholesterol, cholesterol, plasma cholesterol, reduce, protect, increase, independent, no change, mouse, rabbit, human, primate, rat.

Plaque size example: plaque size, plaque area, lesion size, lesion area, reduce, protect, increase, independent, no change, mouse, rabbit, human, primate, rat.

The included results are not meant to be complete, but rather give the reader a simplified toy example with how the proposed metric could be used to ascertain the practical equivalence of results. The mean and standard deviation of each group were either copied directly from the source publication or estimated from the figure using Web Plot Digitizer (https://automeris.io/WebPlotDigitizer/).

*Cases of Dependence Between Changes to Null Strength Measures in Risk Assessment*

We need to test comparison error of statistics across population configurations with changes in value to each null strength measure. In principle, we would alter the independent measure across configurations and hold all other measures constant. Yet this approach is not always possible because the raw and relative null strength measures sometimes covary with each other. For instance, changing $|\mu_{DM}|$ across configurations must also change one or more of $\{|r\mu_{DM}|, r\sigma_{DM}, \sigma_{DM}\}$. This dependence between measures could introduce confounding relationships and prevent us from testing each measure in isolation. If confounding relationships are not dealt with, we cannot conclude if a candidate statistic can determine higher null strength for each null strength measure. For example, if a candidate statistic performs impressively in detecting changes to $|\mu_{DM}|$, we cannot conclude that its performance is due to changes with $|\mu_{DM}|$ since the statistic could be responding from indirect changes to $|r\mu_{DM}|$, $|r\sigma_{DM}|$, or $\sigma_{DM}$.

We avoid this confounding issue by generating sets of population configurations where the ground truth designations between the independent measure and other measures do not have any correlation. While the value of covarying measures may correlate, the ground truth designations can remain uncorrelated from each other with carefully curated population parameter datasets. To accomplish this, we varied the independent measure so that both experiments have an equal and random chance to have higher null strength (50/50 chance). To avoid correlation with ground truth designations from other measures, we generate configurations where the other null strength measures must either:

1) Designate experiment 1 the winner in all cases.
2) Designate experiment 2 the winner in all cases.
3) Designate experiment 1 the winner half the time, but these designations are also random and not correlated with the designations from the independent measure.

All three cases will guarantee that there is no correlation between the ground truth designations of the independent measure and other measures. This lack of correlation is directly verified for each investigation with a binomial test ($H_0$: $\pi$=0.5) of the number of shared ground truth designations between the independent measure and each of the other measures. For an example, Fig S7A visualizes the lack of correlation of the ground truth designations for $|\mu_{DM}|$ compared to the other null strength measures.

### *Parameter Space for Population Configurations in Risk Assessment*

The population configurations were chosen to adequately sample the parameter space to ensure the error rates reflected general trends. A natural way to separate the parameter space is using the previous gold standard of quantifying the statistical significance of results with p-values or some other statistic. Since $\delta_M$ and $r\delta_M$ can be calculated and used regardless of statistical significance, it is not clear if their error rates would be consistent when analyzing null and positive results. We divided the population configurations associated with null and positive results into separate investigations (differing error rates were indeed observed between these two cases in Fig 3).

One method to evaluate statistical significance is to check if the sample t-statistic for $\bar{x}_{DM}$ is less than the critical t-value across the samples generated from the population configurations, where

$$t_{statistic} = \frac{\bar{y} - \bar{x}}{\sqrt{\frac{s_X^2}{m} + \frac{s_Y^2}{n}}} \tag{S1}$$

$$t_{critical} := t_{\alpha_{DM},\; m+n-1} \;. \tag{S2}$$

A result is deemed null if $t_{statistic} < |t_{critical}|$ and positive otherwise (absolute value is used because the distance from zero in either direction is relevant). To summarize the entire collection of samples drawn from a single population parameter, we can compute the mean value of the t-statistic and check if $\bar{t}_{statistic} < |t_{critical}|$. Population configurations where this expression is true is designated as a null configuration. This calculation allows us to separate population configurations associated with null results from positive results, but we also need to separate results based on their degree of statistical significance within each region. We therefore compute the ratio of $\bar{t}_{statistic} / |t_{critical}|$ (we define as the t-ratio) as a method to score results on their statistical significance. Population configurations were chosen to provide wide coverage of null results (absolute expected t-ratio <= 1) and positive results (absolute expected t-ratio > 1). This coverage was typically accomplished by altering the population parameters associated with the independent null strength measure so that there was sufficient coverage. We generated population parameters associated with the independent measure from uniform distributions.

For example, with the investigation where $|\mu_{DM}|$ is the independent measure for null results (SFig. 7), we generated values for $\mu_{DM}$ for experiment 1 and 2 from two different uniform distributions of $U(0.5, 3)$ or $U(2, 4.5)$ (for each population configuration, experiment 1 $\mu_{DM}$ used one of these distributions at random and experiment 2 $\mu_{DM}$ used the other). Two distributions were used so that a large enough difference between $\mu_{DM}$ for experiment 1 and 2 was guaranteed for each population configuration so that candidate statistics would be able to detect the signal in most cases. We wish to minimize cases where the change in null strength between experiment results is so small that none of the candidate statistics can effectively predict which experiment has higher null strength. The standard deviations, sample sizes, and significance levels were set to fixed values between both experiments. While the control mean for the first experiment was $\mu_{X,1} = 20$, the control mean for the second experiment was $\mu_{X,2} = 200$. These values for $\mu_X$ were used so that the null strength measure $|r\mu_{DM}|$ would designate experiment 2 with higher null strength for all population configurations and avoid any correlation between the ground truth

designations for $|r\mu_{DM}|$ and $|\mu_{DM}|$. Please see the associated R script files for more details, and SFig 7B for example histogram of t-ratios of population configurations within the null region.

***Simultaneous Risk Assessment in Risk Assessment***

 Our approach of varying a single measure of null strength at a time for the risk assessment of null strength unfortunately does not simulate real world conditions. It would be reasonable to expect multiple measures of null strength to vary simultaneously when comparing null strength between experiments. To address this shortcoming, we designed population configurations that had multiple null strength measures varied simultaneously as a more realistic scenario.

 We designed a set of population configurations that allowed for all four raw null strength measures to vary simultaneously (Fig 3A, columns under "Simultaneous" header). We examined whether candidate statistics could predict null strength in a better than random fashion by comparing the prediction designations to the ground truth designations for each raw null strength measure. Another set of population configurations were generated that allowed for all relative agreement measures to change simultaneously (Fig 3B, columns under "Simultaneous" header). We examined whether candidate statistics could predict null strength in a better than random fashion by comparing the prediction designations to the ground truth designations for each relative null strength measure.

**A**

$$\partial_M \, Cred. \, Rate = N^{-1} \sum_{i=1}^{K} I\left(\left|\mu_Y^i - \mu_X^i\right| \leq \partial_M\right)$$

**B**

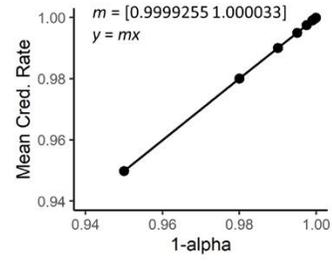

**C**

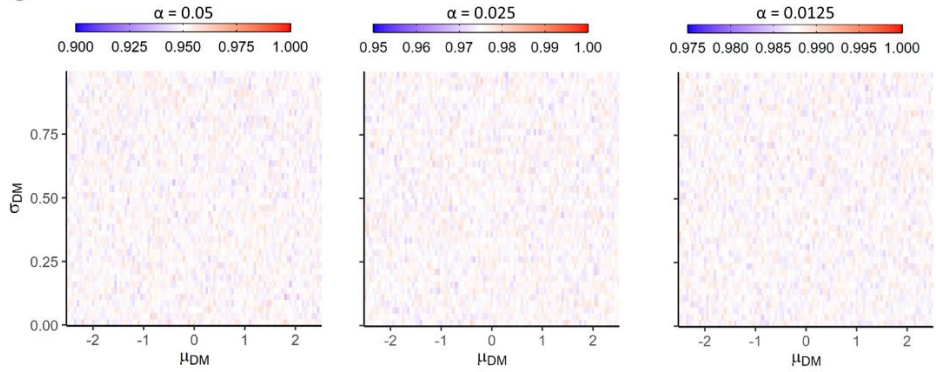

**D**

$$r\partial_M \, Cred. \, Rate = N^{-1} \sum_{i=1}^{K} \left(\frac{\left|\mu_Y^i - \mu_X^i\right|}{\mu_X^i} \leq r\partial_M\right)$$

**E**

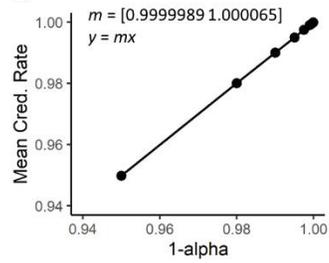

**F**

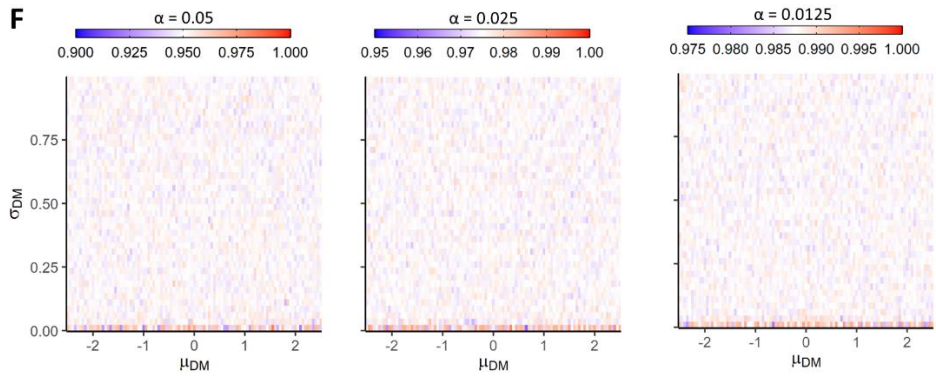

**Fig. S1: Credibility of $\delta_M$ and $r\delta_M$ is controlled by one minus the significance level. (A)** The $\delta_M$ credibility rate is defined as the fraction of Monte Carlo trials simulating the posterior distribution of the difference in means that are less than or equal to $\delta_M$ (with a five-fold increase in trials compared to the calculation for $\delta_M$). **(B)** Mean credibility rate for a range of values for $\overline{x}_{DM}$ and $s_{DM}$ at different significance levels. **(C)** Representative heatmaps of credibility rates at various significance levels. **(D)** The $r\delta_M$ credibility rate is defined as the fraction of Monte Carlo trials simulating the posterior distribution of the relative difference in population means that are less than or equal to $r\delta_M$ (with five-fold increase in trials compared to the calculation for $r\delta_M$). **(E)** Mean credibility for a range of values for $\overline{x}_{DM}$ and $s_{DM}$ at several significance levels. **(F)** Representative heatmaps of credibility rates at various significance levels (equal variances for control and experiment group, N=6 measurements per sample).

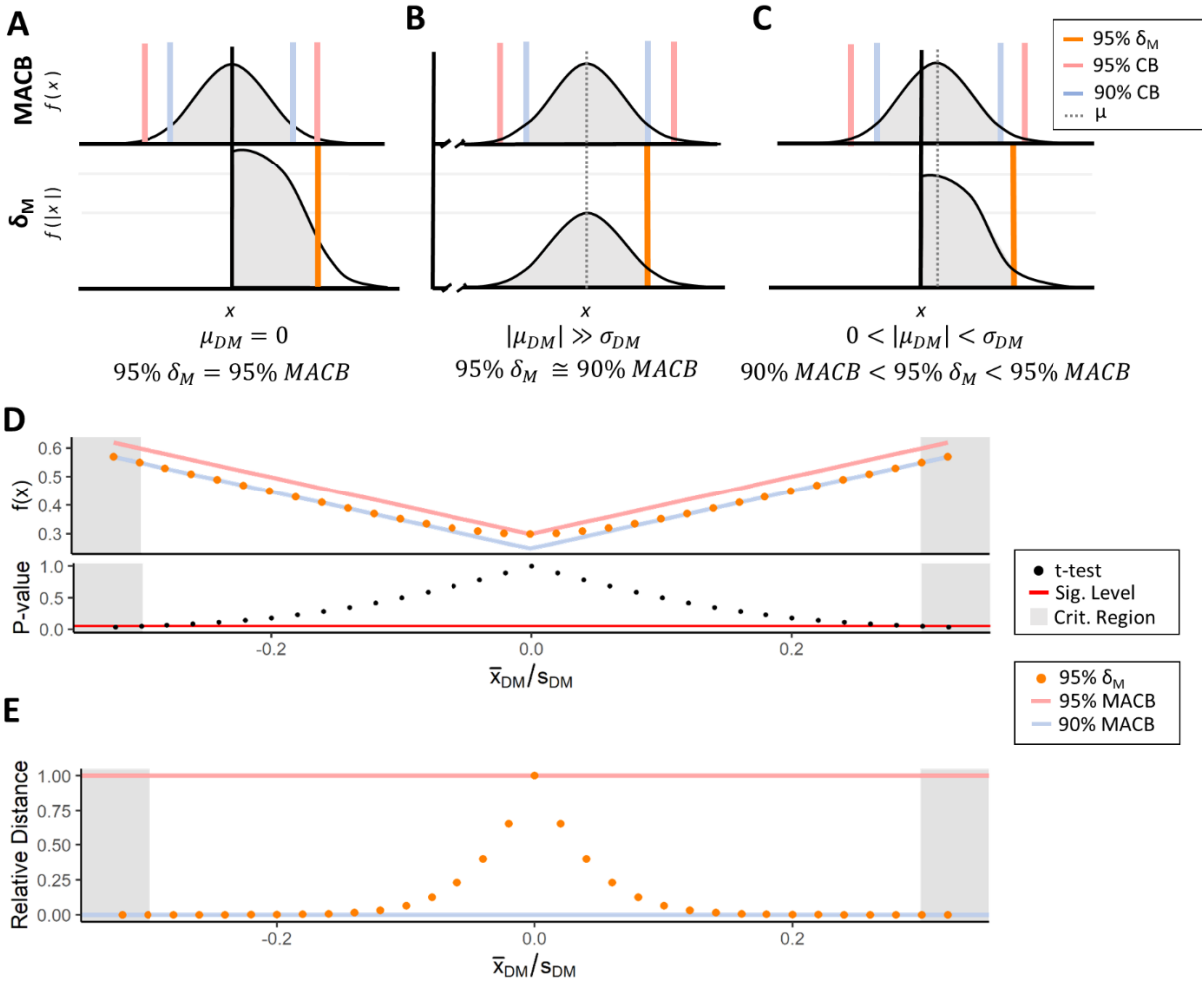

**Fig. S2: Relation between two-tailed credible intervals and δ$_M$.** Graphical depiction of 90% and 95% two-tailed credible interval bounds (CB) of the difference in means, along with the δ$_M$ when the magnitude of μ$_{DM}$ is (**A**) equal to zero, (**B**) much greater than σ$_{DM}$, and (**C**) less than σ$_{DM}$ (with the exact transition point between (B) and (C) unknown). MACB denotes maximum absolute of two-tailed credible bounds of the mean. (**D**) The δ$_M$ and the MACB of a sample shifted over a range of sample means to visualize this transition through the null region (white), with corresponding p-value plotted below (one sample t-test, N=35 observations per point). (**E**) The relative scaling of the 95% δ$_M$ scaled between the max absolute 95% (1.0) and 90% (0.0) credible bounds across a series of normalized samples. Abbreviations: x̄$_{DM}$, sample difference in means; s$_{DM}$, sample standard deviation of the difference in means.

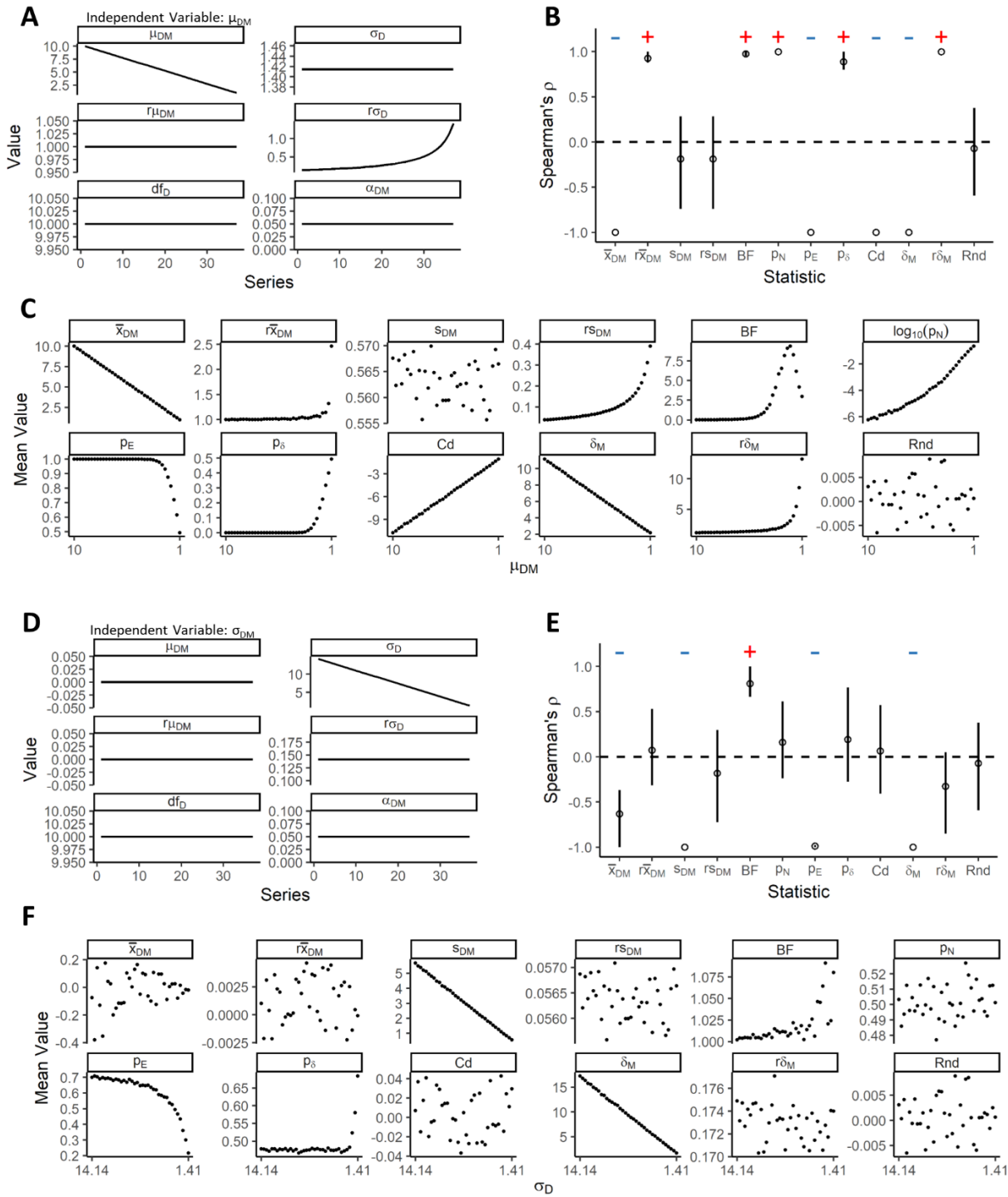

**Fig. S3: correlation of candidate statistics with stronger raw null strength via $\mu_{DM}$ and $\sigma_D$.**
(**A**) A series of population configurations with decreasing $\mu_{DM}$ towards higher null strength with most other null strength measures held constant (changes to $\mu_{DM}$ could not be completely isolated from all other null strength measures, so $r\sigma_D$ also changed with this series, but towards lower null strength). (**B**) Spearman's $\rho$ of each candidate statistic versus $\mu_{DM}$ and (**C**) mean value of candidate statistic across configurations. (**D**) A series of population configurations with decreasing $\sigma_D$ towards higher null strength with all other null strength measures held constant. (**E**) Spearman's $\rho$ of each candidate statistic versus $\sigma_D$ and (**F**) mean value of candidate statistic across configurations. (B, E) Error bars are 95% confidence interval of Spearman's $\rho$ with Bonferroni correction, with red plus denoting candidate statistics with a significant positive correlation and blue minus denoting a significance negative correlation (1E3 samples drawn for each point in the series).

A

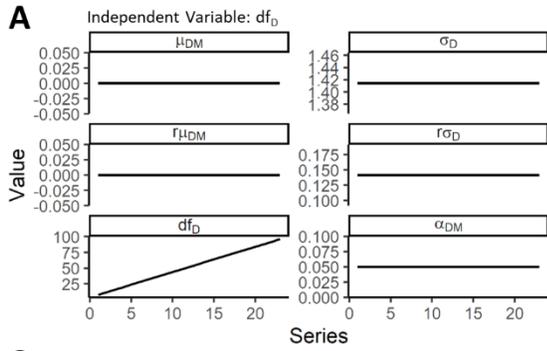

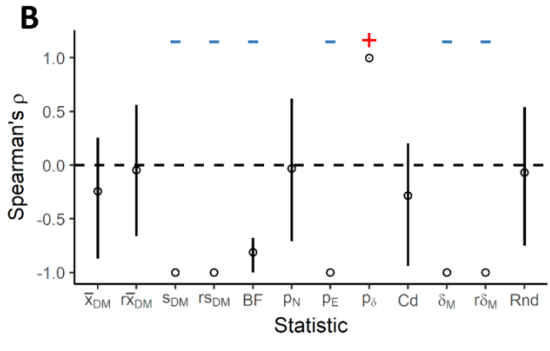

B

C

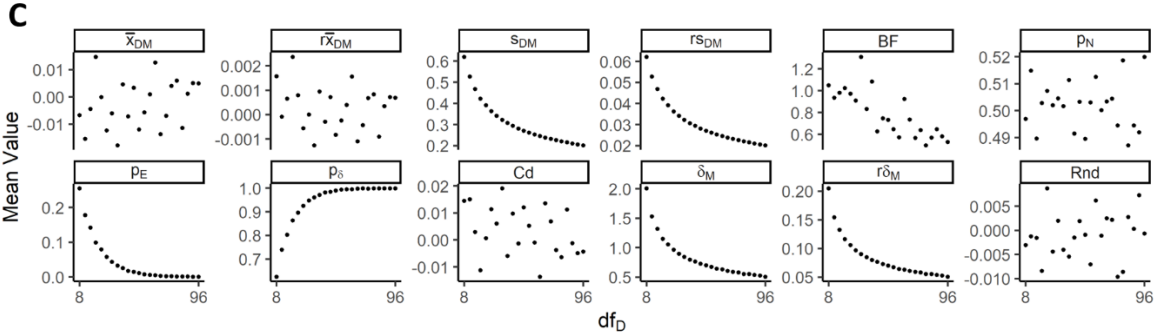

D

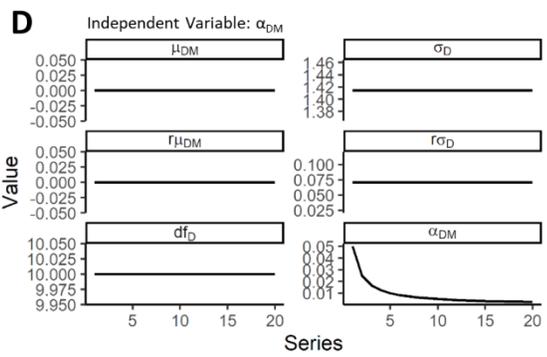

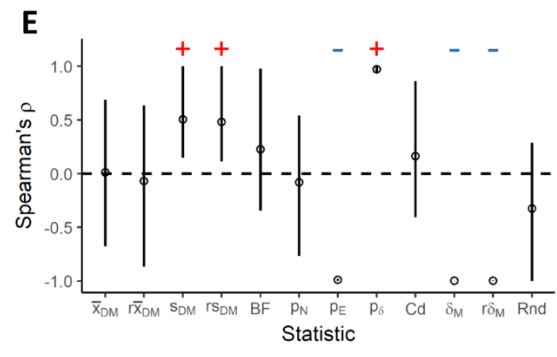

E

F

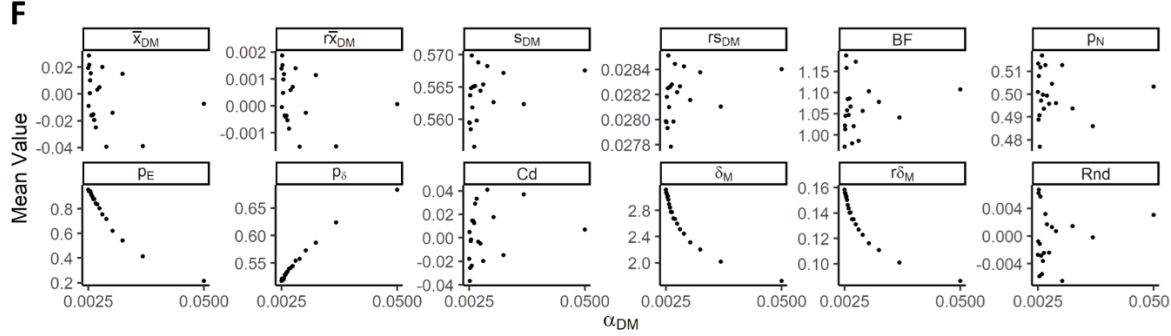

**Fig. S4: correlation of candidate statistics with stronger raw null strength via df$_D$ and α$_{DM}$.**
(**A**) A series of population configurations with decreasing df$_D$ toward higher null strength with all other null strength measures held constant. (**B**) Spearman's ρ of each candidate statistic versus df$_D$ and (**C**) mean value of candidate statistic across configurations. (**D**) A series of population configurations with decreasing α$_{DM}$ toward higher null strength with all other null strength measures held constant. (**E**) Spearman's ρ of each candidate statistic versus α$_{DM}$ and (**F**) mean value of candidate statistic across configurations. (B, E) Error bars are 95% confidence interval of Spearman's ρ with Bonferroni correction, with red plus denoting candidate statistics with a significant positive correlation and blue minus denoting a significance negative correlation.

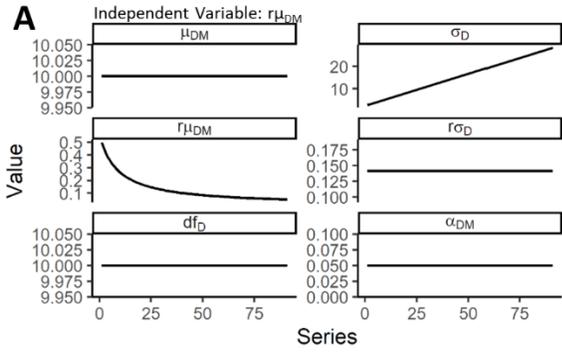

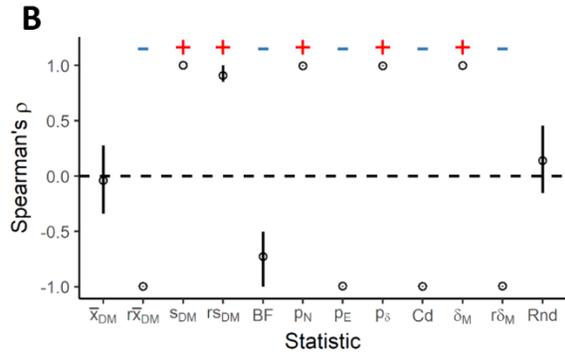

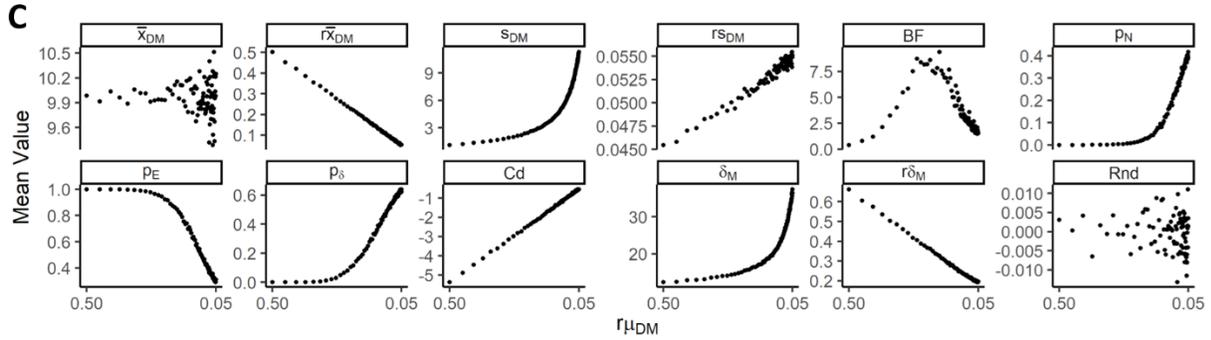

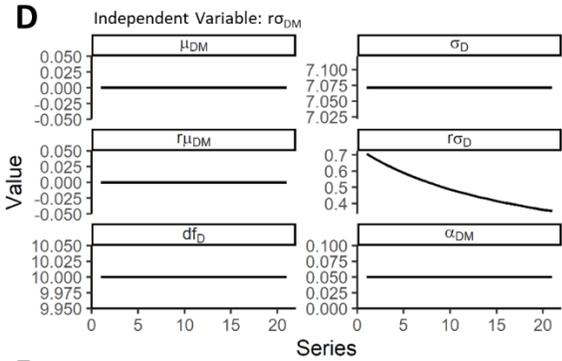

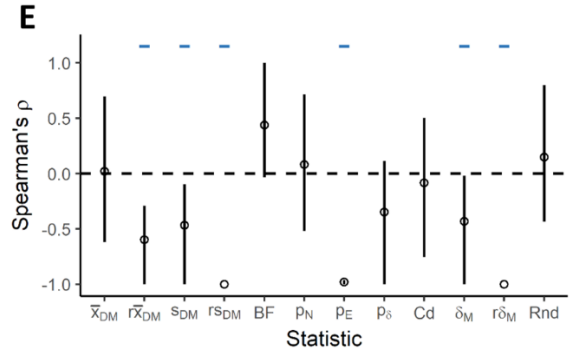

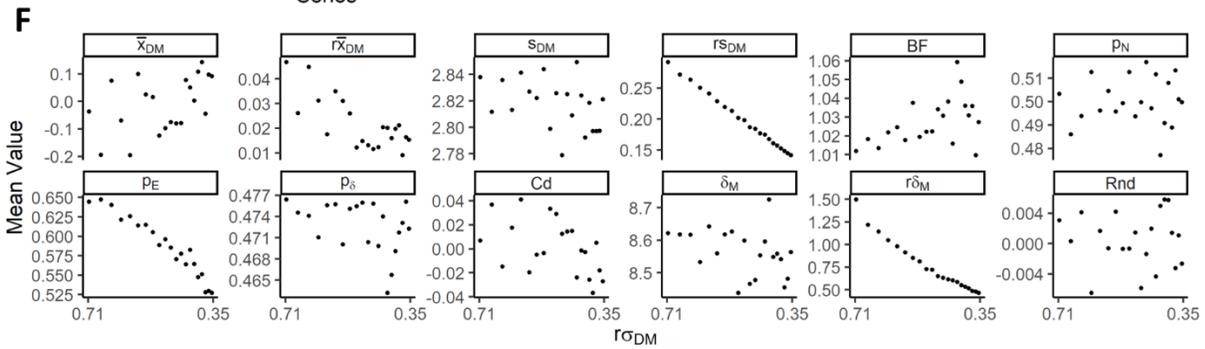

**Fig. S5: correlation of candidate statistics with stronger relative null strength via r$\mu_{DM}$ and r$\sigma_D$.** (**A**) A series of population configurations with decreasing r$\mu_{DM}$ towards higher relative null strength with most other null strength measures held constant (changes to r$\mu_{DM}$ could not be completely isolated from all variables, so $\sigma_D$ also changed with this series, but towards lower null strength). (**B**) Spearman's $\rho$ of each candidate statistic versus r$\mu_{DM}$ and (**C**) mean value of candidate statistic across configurations. (**D**) A series of population configurations with decreasing r$\sigma_D$ towards higher relative null strength with all other null strength measures held constant. (**E**) Spearman's $\rho$ of each candidate statistic versus r$\sigma_D$ and (**F**) mean value of candidate statistic across configurations. (B, E) Error bars are 95% confidence interval of Spearman's $\rho$ with Bonferroni correction, with red plus denoting candidate statistics with a significant positive correlation and blue minus denoting a significance negative correlation (1E3 samples drawn for each point in the series).

**A**

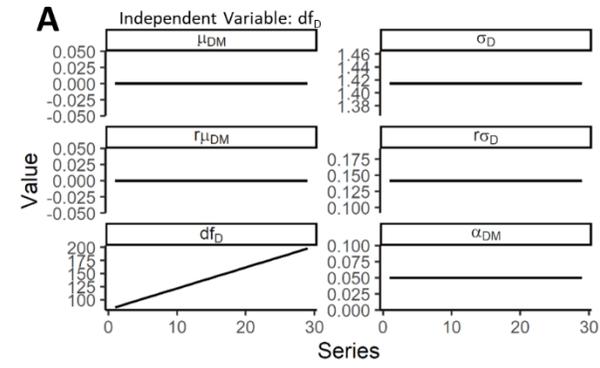

**B**

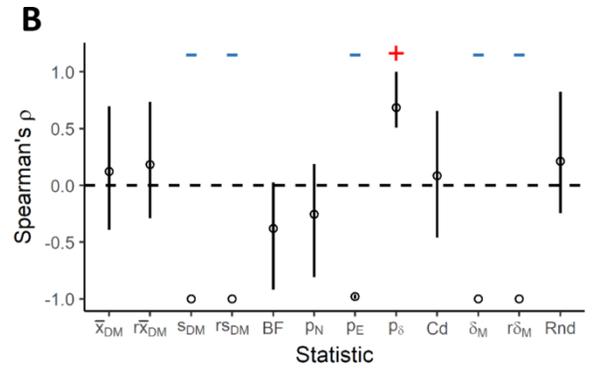

**C**

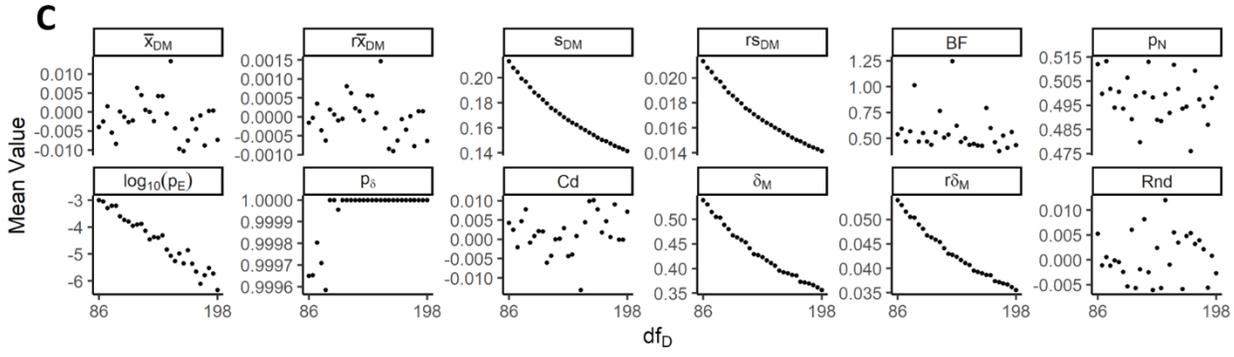

**D**

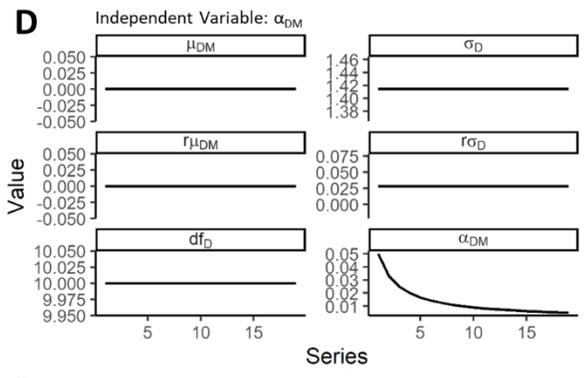

**E**

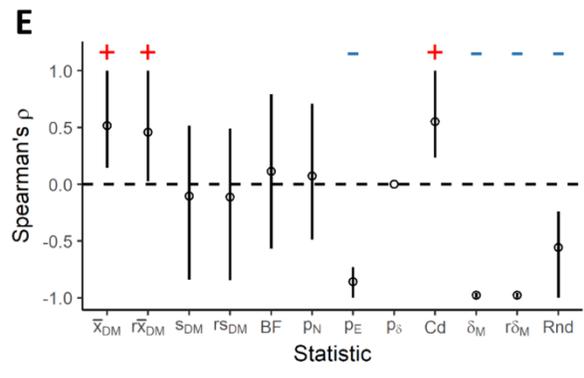

**F**

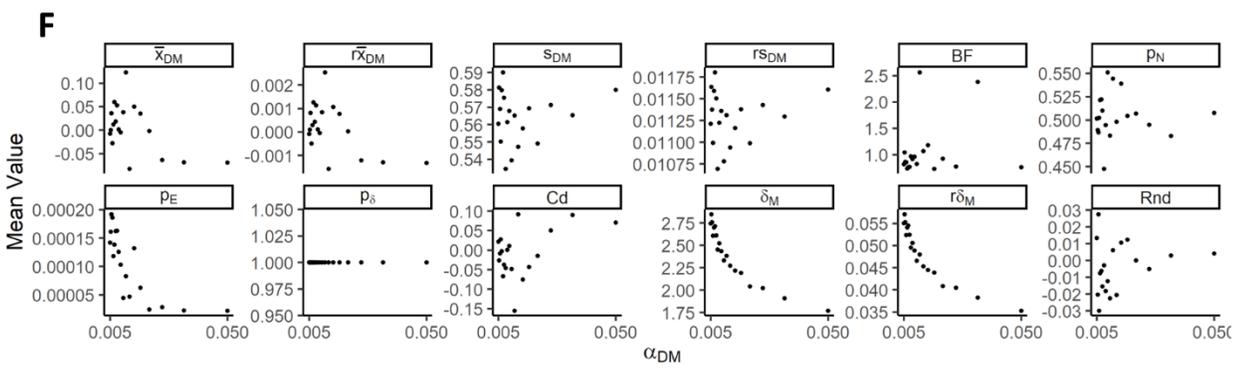

**Fig. S6: correlation of candidate statistics with stronger relative null strength via df$_D$ and α$_{DM}$.** (**A**) A series of population configurations with decreasing df$_D$ toward higher relative null strength with all other null strength measures held constant. (**B**) Spearman's ρ of each candidate statistic versus df$_D$ and (**C**) mean value of candidate statistic across configurations. (**D**) A series of population configurations with α$_{DM}$ reduced toward higher relative null strength with all other null strength measures held constant. (**E**) Spearman's ρ of each candidate statistic versus α$_{DM}$ and (**F**) mean value of candidate statistic across configurations. (B, E) Error bars are 95% confidence interval of Spearman's ρ with Bonferroni correction, with red plus denoting candidate statistics with a significant positive correlation and blue minus denoting a significant negative correlation (1E3 samples drawn for each point in the series).

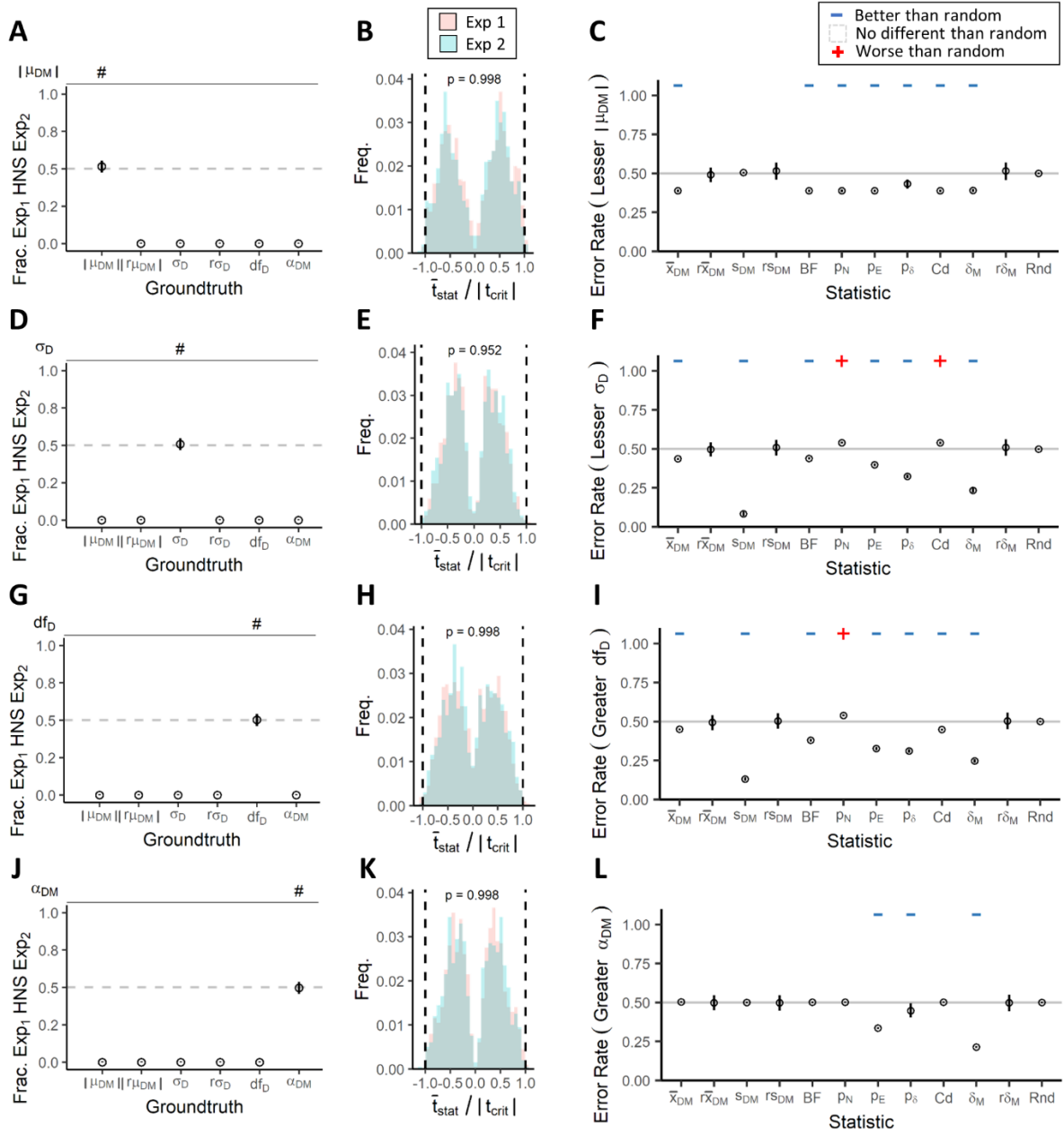

**Fig. S7: The $\delta_M$ is the only statistic that has lower than random comparison error with null results for each measure of null strength.** (**A**) Fraction of population configurations where experiment 1 has higher null strength than (HNS) experiment 2 according to each measure of null strength, with designations from $\mu_{DM}$ serving as ground truth. (**B**) Histogram of the ratio of $\overline{t_{statistic}}$ *to* $t_{critical}$ for population configurations, indicating that results from experiment 1 (blue) and experiment 2 (pink) are both associated with null results ($|\overline{t_{statistic}} / t_{critical}| < 1$). (**C**) Mean comparison error rate of candidate statistics in identifying which experiment has higher null strength via lower $\mu_{DM}$ across population configurations (50 observations per sample). (**D**) Fraction of population configurations where experiment 1 has higher null strength than experiment 2 according to each measure of null strength, with designations from $\sigma_D$ serving as ground truth. (**E**) Histogram of the ratio of $\overline{t_{statistic}}$ to $t_{critical}$ indicating population configurations are associated with null results. (**F**) Mean comparison error rate of candidate statistics in identifying which experiment has higher null strength via lower $\sigma_D$ across population configurations (50 observations per sample). (**G**) Fraction of population configurations where experiment 1 has higher null strength than experiment 2 according to each measure of null strength, with designations from $df_D$ serving as ground truth. (**H**) Histogram of the ratio of $\overline{t_{statistic}}$ *to* $t_{critical}$ indicating population configurations are associated with null results. (**I**) Mean comparison error rate of candidate statistics in identifying which experiment has higher null strength via higher $df_D$ across population configurations (6 - 40 observations per sample). (**J**) Fraction of population configurations where experiment 1 has higher null strength than experiment 2 according to each measure of null strength, with designations from $\alpha_{DM}$ serving as ground truth. (**K**) Histogram of the ratio of $\overline{t_{statistic}}$ *to* $t_{critical}$ indicating population configurations are associated with null results. (**L**) Mean comparison error rate of candidate statistics in identifying which experiment has higher null strength via higher $\alpha_{DM}$ across population configurations (30 observations per sample). (A, D, G, J) '#' denotes measures that have a nonrandom number of shared designations with independent measures (listed at top of y-axis) for which experiments are designated with higher null strength ($p < 0.05$ from Bonferroni corrected two-tailed binomial test for coefficient equal to 0.5 between independent measure and each null strength measure). (B, E, H, K) Discrete Kolmogorov-Smirnov test between histograms. (C, F, I, L) Pairwise t-test with Bonferroni correction for all combinations, where blue minus denotes a mean error rate lower than random, red plus denotes higher than random. N=1E3 population configurations generated for each study, n=1E2 samples drawn per configuration.

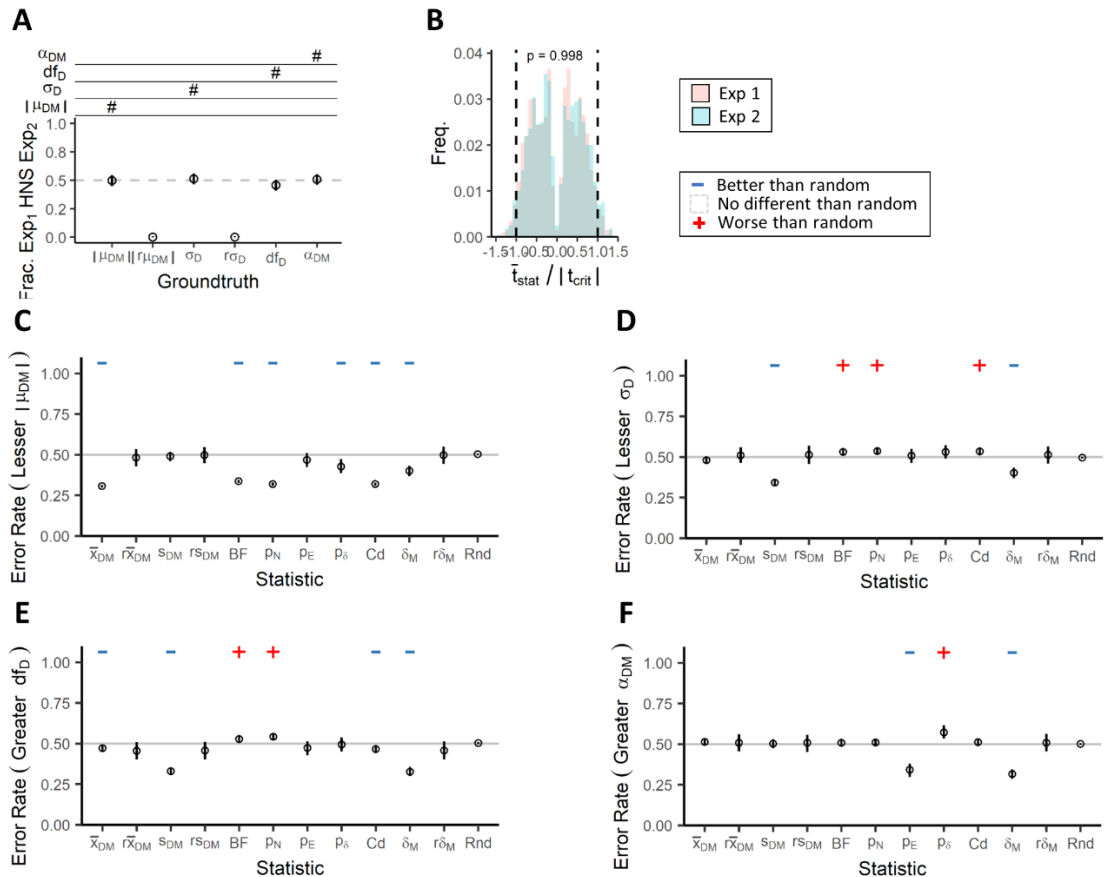

**Fig. S8: The δ<sub>M</sub> is the only statistic that has lower than random comparison error with null results across all measures of null strength simultaneously.** (**A**) Fraction of population configurations where experiment 1 has higher null strength than (HNS) experiment 2 according to each measure of null strength, with designations from $\mu_{DM}$, $\sigma_D$, $df_D$, and $\alpha_{DM}$ serving as separate ground truths simultaneously. (**B**) Histogram of the ratio of $\bar{t}_{statistic}$ *to* $t_{critical}$ for population configurations, indicating that results from experiment 1 (blue) and experiment 2 (pink) are both associated with null results ($|\bar{t}_{statistic} / t_{critical}| < 1$). From a single data set, mean comparison error rate of candidate statistics in identifying which experiment has higher null strength via (**C**) lower $\mu_{DM}$, (**D**) lower $\sigma_D$, (**E**) higher $df_D$, and (**F**) higher $\alpha_{DM}$ across population configurations. (A) '#' denotes measures that have a nonrandom number of shared designations with each independent measure (listed at top of y-axis) for which experiments are designated with higher null strength ($p < 0.05$ from Bonferroni corrected two-tailed binomial test for coefficient equal to 0.5 between each independent measure and every null strength measure). (B) Discrete Kolmogorov-Smirnov test between histograms. (C, D, E, F) Pairwise t-test with Bonferroni correction for all combinations, where blue minus denotes a mean error rate lower than random, red plus denotes higher than random. N=1E3 population configurations generated for each study, n=1E2 samples drawn per configuration, 5 - 20 observations per sample.

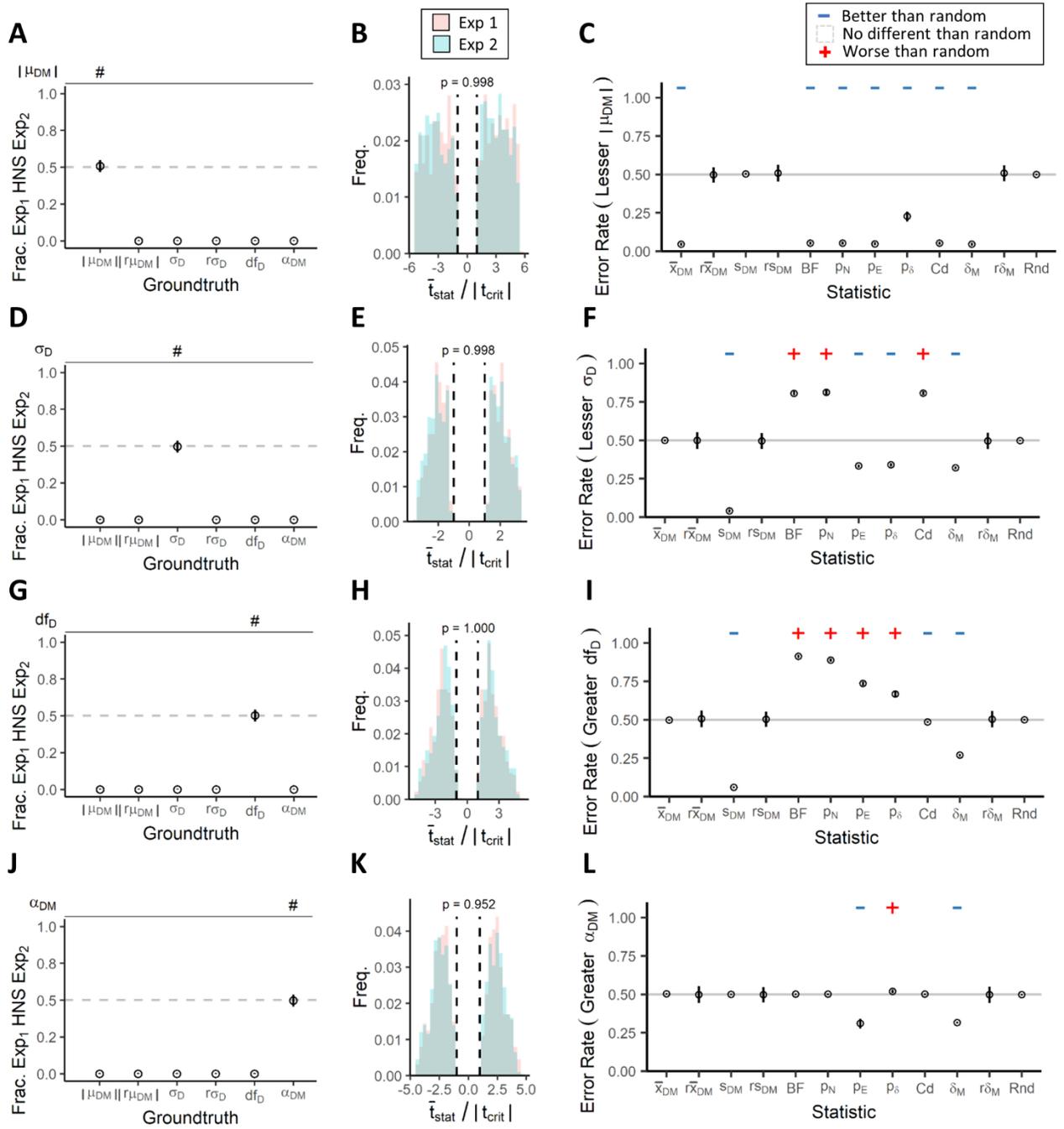

**Fig. S9: The $\delta_M$ is the only statistic that has lower than random comparison error with positive results for each measure of null strength.** (**A**) Fraction of population configurations where experiment 1 has higher null strength than (HNS) experiment 2 according to each measure of null strength, with designations from $\mu_{DM}$ serving as ground truth. (**B**) Histogram of the ratio of $\bar{t}_{statistic}$ to $t_{critical}$ for population configurations, indicating that results from experiment 1 (blue) and experiment 2 (pink) are both associated with positive results ($|\bar{t}_{statistic} / t_{critical}| > 1$). (**C**) Mean comparison error rate of candidate statistics in identifying which experiment has higher null strength via lower $\mu_{DM}$ across population configurations (50 observations per sample). (**D**) Fraction of population configurations where experiment 1 has higher null strength than experiment 2 according to each measure of null strength, with designations from $\sigma_D$ serving as ground truth. (**E**) Histogram of the ratio of $\bar{t}_{statistic}$ to $t_{critical}$ indicating that population configurations are associated with positive results. (**F**) Mean comparison error rate of candidate statistics in identifying which experiment has higher null strength via lower $\sigma_D$ across population configurations (50 observations per sample). (**G**) Fraction of population configurations where experiment 1 has higher null strength than experiment 2 according to each measure of null strength, with designations from $df_D$ serving as ground truth. (**H**) Histogram of the ratio of $\bar{t}_{statistic}$ to $t_{critical}$ indicating that population configurations are associated with positive results. (**I**) Mean comparison error rate of candidate statistics in identifying which experiment has higher null strength via higher $df_D$ across population configurations (6 - 40 observations per sample). (**J**) Fraction of population configurations where experiment 1 has higher null strength than experiment 2 according to each measure of null strength, with designations from $\alpha_{DM}$ serving as ground truth. (**K**) Histogram of the ratio of $\bar{t}_{statistic}$ to $t_{critical}$ indicating that population configurations are associated with positive results. (**L**) Mean comparison error rate of candidate statistics in identifying which experiment has higher null strength via higher $\alpha_{DM}$ across population configurations (30 observations per sample). (A, D, G, J) '#' denotes measures that have a nonrandom number of shared designations with independent measure (listed at top of y-axis) for which experiments are designated with higher null strength ($p < 0.05$ from Bonferroni corrected two-tailed binomial test for coefficient equal to 0.5 between independent measure and each null strength measure). (B, E, H, K) Discrete Kolmogorov-Smirnov test between histograms. (C, F, I, L) Pairwise t-test with Bonferroni correction for all combinations, where blue minus denotes a mean error rate lower than random, red plus denotes higher than random. N=1E3 population configurations generated for each study, n=1E2 samples drawn per configuration.

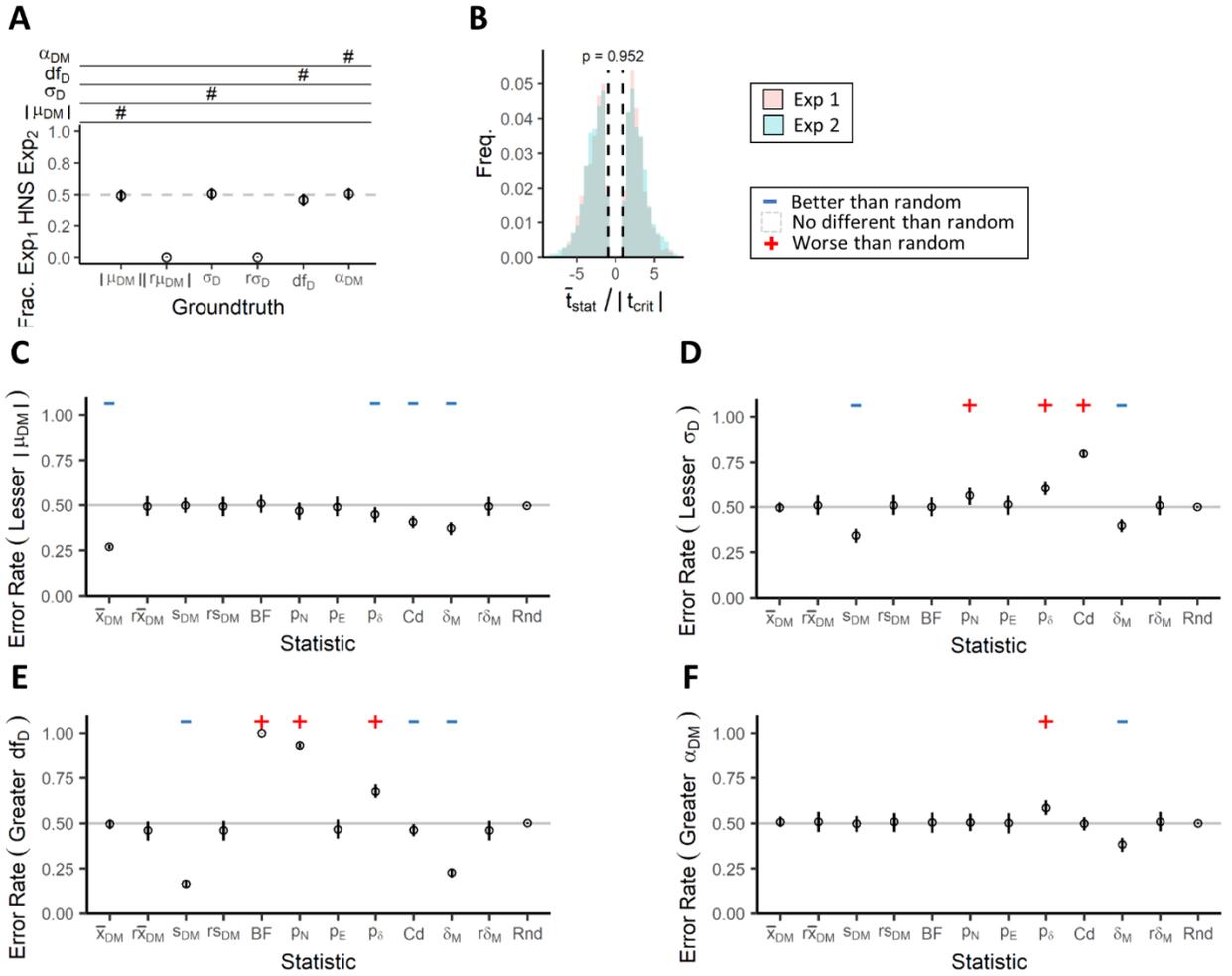

**Fig. S10: The δ_M is the only statistic that has lower than random comparison error with positive results across all measures of null strength simultaneously.** (**A**) Fraction of population configurations where experiment 1 has higher null strength than (HNS) experiment 2 according to each measure of null strength, with designations from $\mu_{DM}$, $\sigma_D$, $df_D$, and $\alpha_{DM}$ serving as separate ground truths simultaneously. (**B**) Histogram of the ratio of $\bar{t}_{statistic}$ *to* $t_{critical}$ for population configurations, indicating that results from experiment 1 (blue) and experiment 2 (pink) are both associated with positive results ($|\bar{t}_{statistic} / t_{critical}| > 1$). From a single data set, mean comparison error rate of candidate statistics in identifying which experiment has higher null strength via (**C**) lower $\mu_{DM}$, (**D**) lower $\sigma_D$, (**E**) higher $df_D$, and (**F**) higher $\alpha_{DM}$ across population configurations. (A) '#' denotes measures that have a nonrandom number of shared designations with each independent measure (listed at top of y-axis) for which experiments are designated with higher null strength ($p < 0.05$ from Bonferroni corrected two-tailed binomial test for coefficient equal to 0.5 between each independent measure and every null strength measure). (B) Discrete Kolmogorov-Smirnov test between histograms. (C, D, E, F) Pairwise t-test with Bonferroni correction for all combinations, where blue minus denotes a mean error rate lower than random, red plus denotes higher than random. N=1E3 population configurations generated for each study, n=1E2 samples drawn per configuration, 6 - 30 observations per sample.

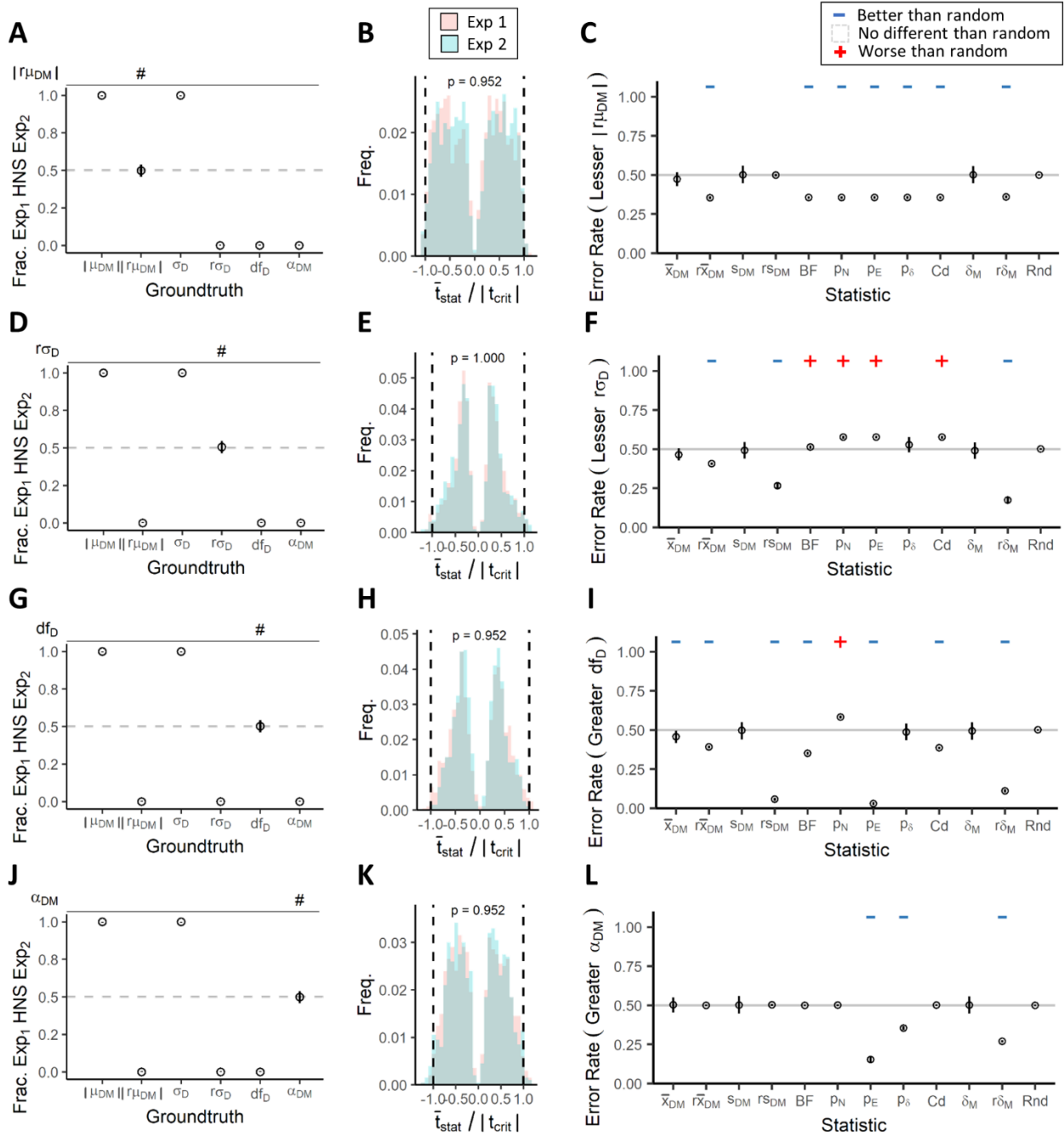

**Fig. S11: The r$\delta_M$ is the only statistic that has lower than random comparison error with null results for each measure of relative null strength.** (**A**) Fraction of population configurations where experiment 1 has higher null strength than (HNS) experiment 2 according to each measure of null strength, with designations from r$\mu_{DM}$ serving as ground truth. (**B**) Histogram of the ratio of $\overline{t_{statistic}}$ to $t_{critical}$ for population configurations, indicating that results from experiment 1 (blue) and experiment 2 (pink) are both associated with null results ($\mu_{DM}/\sigma_{DM}$ < 2.5). (**C**) Mean comparison error rate of candidate statistics in identifying which experiment has lower relative null strength via lower r$\mu_{DM}$ across population configurations (50 observations per sample). (**D**) Fraction of population configurations where experiment 1 has higher null strength than experiment 2 according to each measure of null strength, with designations from r$\sigma_D$ serving as ground truth. (**E**) Histogram of the ratio of $\overline{t_{statistic}}$ to $t_{critical}$ indicating population configurations are associated with null results. (**F**) Mean comparison error rate of candidate statistics in identifying which experiment has lower relative null strength via lower r$\sigma_D$ across population configurations (50 observations per sample). (**G**) Fraction of population configurations where experiment 1 has higher null strength than experiment 2 according to each measure of null strength, with designations from df$_D$ serving as ground truth. (**H**) Histogram of the ratio of $\overline{t_{statistic}}$ to $t_{critical}$ indicating population configurations are associated with null results. (**I**) Mean comparison error rate of candidate statistics in identifying which experiment has lower relative null strength via higher df$_D$ across population configurations (6 - 30 observations per sample). (**J**) Fraction of population configurations where experiment 1 has higher null strength than experiment 2 according to each measure of null strength, with designations from $\alpha_{DM}$ serving as ground truth. (**K**) Histogram of the ratio of $\overline{t_{statistic}}$ to $t_{critical}$ indicating population configurations are associated with null results. (**L**) Mean comparison error rate of candidate statistics in identifying which experiment has lower relative null strength via higher $\alpha_{DM}$ across population configurations (30 observations per sample). (A, D, G, J) '#' denotes measures that have a nonrandom number of shared designations with independent measure (listed at top of y-axis) for which experiments are designated with higher null strength (p < 0.05 from Bonferroni corrected two-tailed binomial test for coefficient equal to 0.5 between independent measure and each null strength measure). (B, E, H, K) Discrete Kolmogorov-Smirnov test between histograms. (C, F, I, L) Pairwise t-test with Bonferroni correction for all combinations, where blue minus denotes a mean error rate lower than random, red plus denotes higher than random. N=1E3 population configurations generated for each study, n=1E2 samples drawn per configuration.

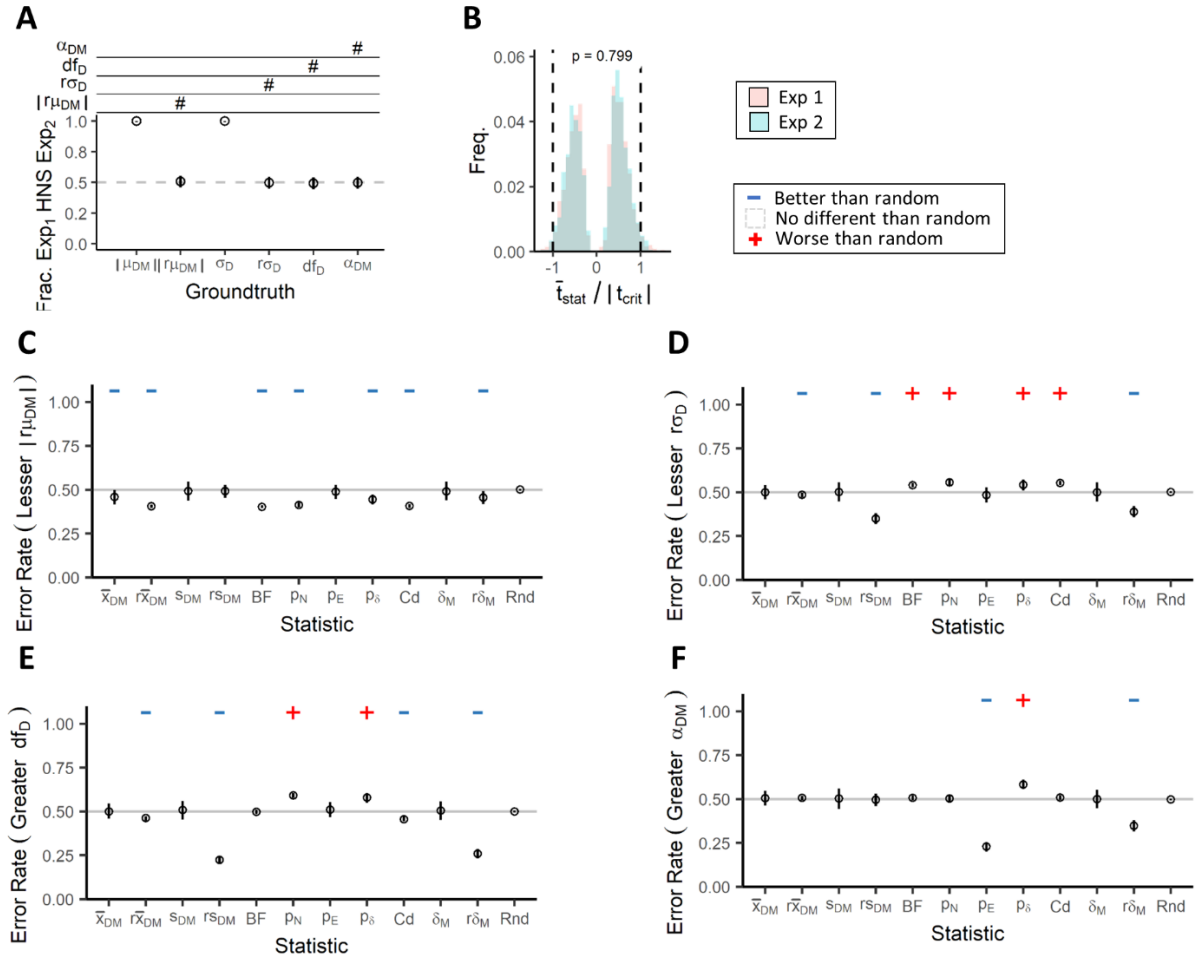

**Fig. S12: The rδ_M is the only statistic that has lower than random comparison error with null results across all measures of relative null strength simultaneously.** (**A**) Fraction of population configurations where experiment 1 has higher null strength than (HNS) experiment 2 according to each measure of null strength, with designations from rμ_DM, rσ_D, df_D, and α_DM serving as separate ground truths simultaneously. (**B**) Histogram of the ratio of $\bar{t}_{statistic}$ to $t_{critical}$ for population configurations, indicating that results from experiment 1 (blue) and experiment 2 (pink) are both associated with null results ($|\bar{t}_{statistic} / t_{critical}| \leq 1$). From a single data set, mean comparison error rate of candidate statistics in identifying which experiment has lower relative null strength via (**C**) lower rμ_DM, (**D**) lower rσ_D, (**E**) higher df_D, and (**F**) higher α_DM across population configurations. (A) '#' denotes measures that have a nonrandom number of shared designations with each independent measure (listed at top of y-axis) for which experiments are designated with higher null strength ($p < 0.05$ from Bonferroni corrected two-tailed binomial test for coefficient equal to 0.5 between each independent measure and every null strength measure). (B) Discrete Kolmogorov-Smirnov test between histograms. (C, D, E, F) Pairwise t-test with Bonferroni correction for all combinations, where blue minus denotes a mean error rate lower than random, red plus denotes higher than random. N=1E3 population configurations generated for each study, n=1E2 samples drawn per configuration, 5 - 30 observations per sample.

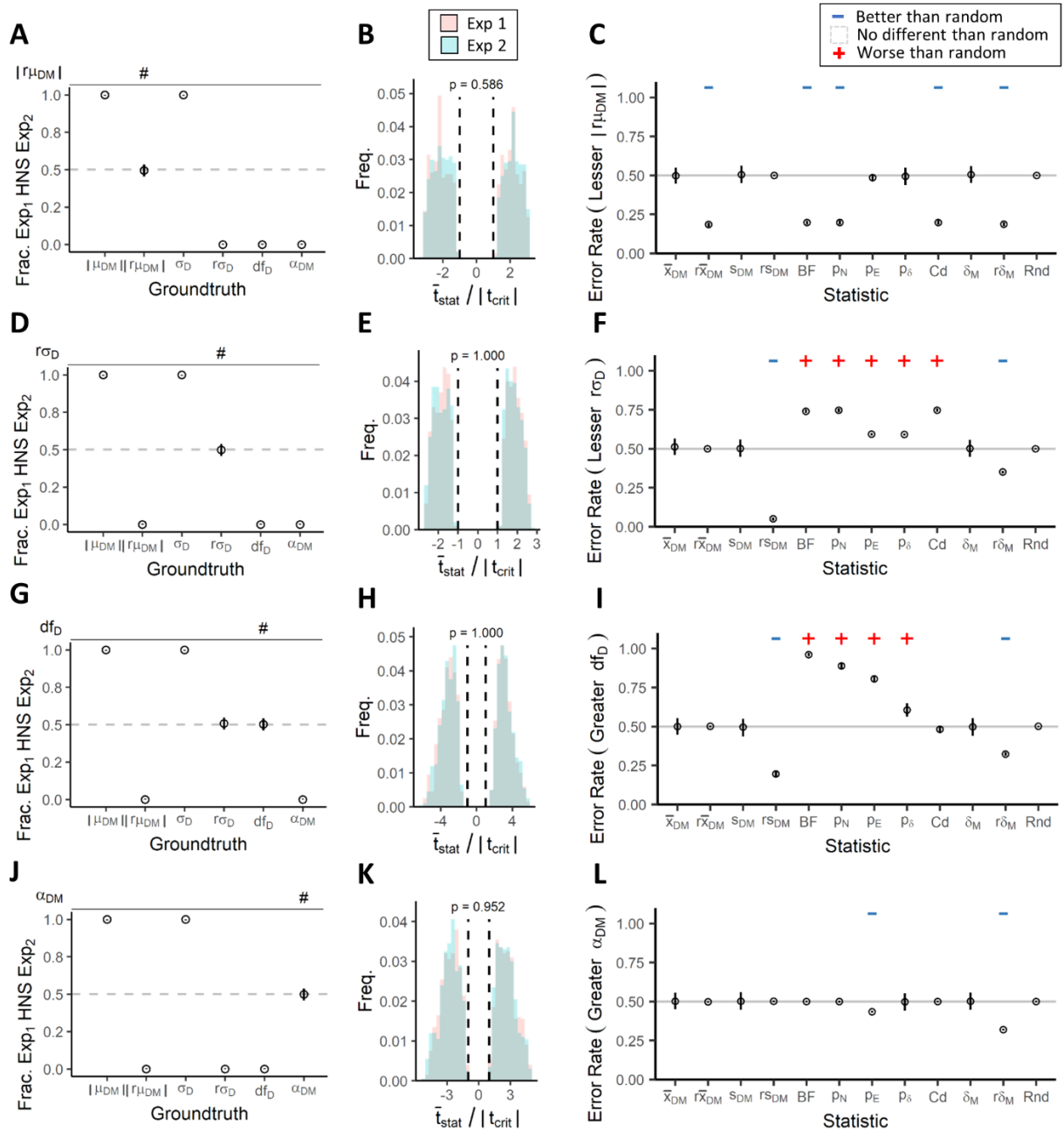

**Fig. S13: The r$\delta_M$ is the only statistic that has lower than random comparison error with positive results for each measure of relative null strength.** (**A**) Fraction of population configurations where experiment 1 has higher null strength than (HNS) experiment 2 according to each measure of null strength, with designations from r$\mu_{DM}$ serving as ground truth. (**B**) Histogram of the ratio of $\bar{t}_{statistic}$ *to* $t_{critical}$ for population configurations, indicating that results from experiment 1 (blue) and experiment 2 (pink) are both associated with positive results ($|\bar{t}_{statistic} / t_{critical}| > 1$). (**C**) Mean comparison error rate of candidate statistics in identifying which experiment has lower relative null strength via lower r$\mu_{DM}$ across population configurations (50 observations per sample). (**D**) Fraction of population configurations where experiment 1 has higher null strength than experiment 2 according to each measure of null strength, with designations from r$\sigma_D$ serving as ground truth. (**E**) Histogram of the ratio of $\bar{t}_{statistic}$ *to* $t_{critical}$ indicating population configurations are associated with positive results. (**F**) Mean comparison error rate of candidate statistics in identifying which experiment has lower relative null strength via lower r$\sigma_D$ across population configurations (50 observations per sample). (**G**) Fraction of population configurations where experiment 1 has higher null strength than experiment 2 according to each measure of null strength, with designations from df$_D$ serving as ground truth. (**H**) Histogram of the ratio of $\bar{t}_{statistic}$ *to* $t_{critical}$ indicating population configurations are associated with positive results. (**I**) Mean comparison error rate of candidate statistics in identifying which experiment has lower relative null strength via higher df$_D$ across population configurations (6 - 30 observations per sample). (**J**) Fraction of population configurations where experiment 1 has higher null strength than experiment 2 according to each measure of null strength, with designations from $\alpha_{DM}$ serving as ground truth. (**K**) Histogram of the ratio of $\bar{t}_{statistic}$ *to* $t_{critical}$ indicating population configurations are associated with positive results. (**L**) Mean comparison error rate of candidate statistics in identifying which experiment has lower relative null strength via higher $\alpha_{DM}$ across population configurations (50 observations per sample). (A, D, G, J) '#' denotes measures that have a nonrandom number of shared designations with independent measure (listed at top of y-axis) for which experiments are designated with higher null strength ($p < 0.05$ from Bonferroni corrected two-tailed binomial test for coefficient equal to 0.5 between independent measure and each null strength measure). (B, E, H, K) Discrete Kolmogorov-Smirnov test between histograms. (C, F, I, L) Pairwise t-test with Bonferroni correction for all combinations, where blue minus denotes a mean error rate lower than random, red plus denotes higher than random. N=1E3 population configurations generated for each study, n=1E2 samples drawn per configuration.

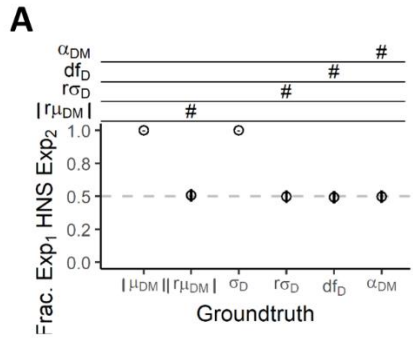

**A**

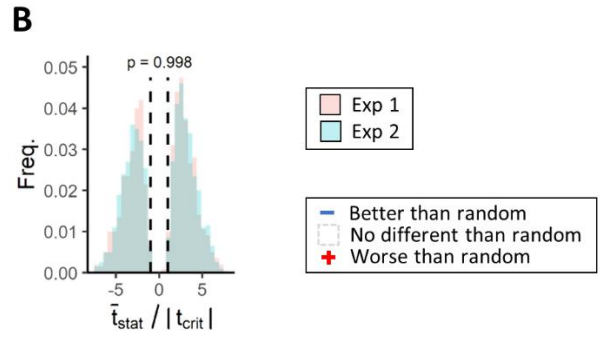

**B**

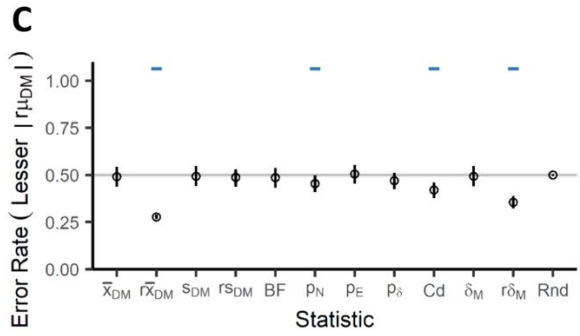

**C**

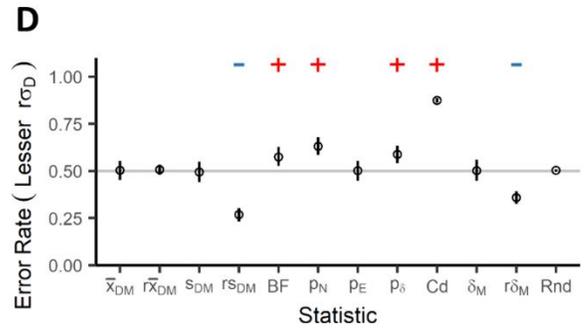

**D**

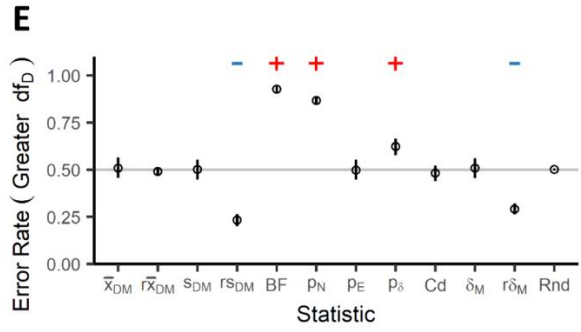

**E**

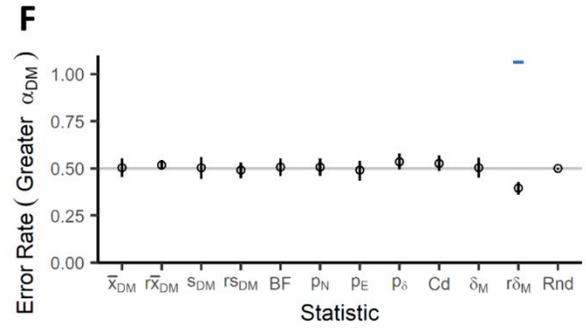

**F**

**Fig. S14: The r$\delta_M$ is the only statistic that has lower than random comparison error with positive results across all measures of relative null strength simultaneously.** (**A**) Fraction of population configurations where experiment 1 has higher null strength than (HNS) experiment 2 according to each measure of null strength, with designations from r$\mu_{DM}$, r$\sigma_D$, df$_D$, and $\alpha_{DM}$ serving as separate ground truths simultaneously. (**B**) Histogram of the ratio of $\bar{t}_{statistic}$ to $t_{critical}$ for population configurations, indicating that results from experiment 1 (blue) and experiment 2 (pink) are both associated with positive results ($|\bar{t}_{statistic} / t_{critical}| > 1$). From a single data set, mean comparison error rate of candidate statistics in identifying which experiment has lower relative null strength via (**C**) lower r$\mu_{DM}$, (**D**) lower r$\sigma_D$, (**E**) higher df$_D$, and (**F**) higher $\alpha_{DM}$ across population configurations. (A) '#' denotes measures that have a nonrandom number of shared designations with each independent measure (listed at top of y-axis) for which experiments are designated with higher null strength (p < 0.05 from Bonferroni corrected two-tailed binomial test for coefficient equal to 0.5 between each independent measure and every null strength measure). (B) Discrete Kolmogorov-Smirnov test between histograms. (C, D, E, F) Pairwise t-test with Bonferroni correction for all combinations, where blue minus denotes a mean error rate lower than random, red plus denotes higher than random. N=1E3 population configurations generated for each study, n=1E2 samples drawn per configuration, 6 - 50 observations per sample.

**Table S1:** Candidate Summary Statistics to Evaluate Comparison Error

| Statistic | Equation | Decision Rule |
|---|---|---|
| $\bar{x}_{\text{DM}}$ | $\bar{y} - \bar{x}$ | $|\bar{x}_{\text{DM},1}| < |\bar{x}_{\text{DM},2}|$ |
| $r\bar{x}_{\text{DM}}$ | $\dfrac{\bar{x}_{DM}}{\bar{x}}$ | $|\bar{x}_{\text{DM},1}| < |\bar{x}_{\text{DM},2}|$ |
| $s_{\text{DM}}$ | $\sqrt{\dfrac{s_X^2}{m} + \dfrac{s_Y^2}{n}}$ | $|s_{\text{DM},1}| < |s_{\text{DM},2}|$ |
| $rs_{\text{DM}}$ | $\dfrac{s_{DM}}{\bar{x}}$ | $|rs_{\text{DM},1}| < |rs_{\text{DM},2}|$ |
| BF ($1$) | $\dfrac{Pr(D|M_1)}{Pr(D|M_2)}$ | $\text{BF}_1 < \text{BF}_2$ |
| $p_{\text{N}}$ ($2$) | $P\left( Z \geq \dfrac{\bar{y} - \bar{x}}{\sqrt{\sigma_X^2/m + \sigma_Y^2/n}} \right)$ | $p_{\text{NHST},1} > p_{\text{NHST},2}$ |
| $p_{\text{E}}$ ($3$) | $Max\left\{ \left( \Delta \leq \dfrac{\bar{y} - \bar{x} - \Delta_L}{\sqrt{\sigma_X^2/m + \sigma_Y^2/n}} \right), \left( \Delta \geq \dfrac{\bar{y} - \bar{x} + \Delta_U}{\sqrt{\sigma_X^2/m + \sigma_Y^2/n}} \right) \right\}$ | $p_{\text{E},1} > p_{\text{PE},2}$ |
| $p_\delta$ ($4$) | $p_\delta = \dfrac{|I \cap H_0|}{|I|} \times max\left\{ \dfrac{|I|}{2|H_0|}, 1 \right\}$ | $p_{\delta,1} > p_{\delta,2}$ |
| CD ($5$) | $\dfrac{\bar{y} - \bar{x}}{\sqrt{\dfrac{(m-1)s_X^2 + (n-1)s_Y^2}{m+n-2}}}$ | $|\text{CD}_1| < |\text{CD}_2|$ |
| $\delta_{\text{M}}$ | $N^{-1} \displaystyle\sum_{i=1}^{K} \mathbb{I}(\mu_Y^i - \mu_X^i \leq c) - N^{-1} \sum_{i=1}^{K} \mathbb{I}(\mu_Y^i - \mu_X^i \leq -c) = 1 - \alpha_{DM}$ | $\delta_{\text{M},1} < \delta_{\text{M},2}$ |
| $r\hat{\delta}_{\text{M}}$ | $N^{-1} \displaystyle\sum_{i=1}^{K} \mathbb{I}\left( \dfrac{\mu_Y^i - \mu_X^i}{\mu_X^i} \leq c \right) - N^{-1} \sum_{i=1}^{K} \mathbb{I}\left( \dfrac{\mu_Y^i - \mu_X^i}{\mu_X^i} \leq -c \right) = 1 - \alpha_{DM}$ | $r\hat{\delta}_{\text{M},1} < r\hat{\delta}_{\text{M},2}$ |
| Rnd | | $\text{Rnd}_1 < \text{Rnd}_2$ |

*Note: decision rule expression is true if experiment 1 is designated with higher null strength than experiment 2.*

*Note 2: noninformative prior used for BF, confidence interval used for $p_\delta$ since testing was primarily used frequentist approaches.*

*Abbreviations: $\bar{x}$, sample mean of control group; $\bar{y}$, sample mean of experiment group; $s_X$, sample standard deviation of control group; $s_Y$, sample standard deviation of experiment group; $\bar{x}_{DM}$, difference in sample means; $r\bar{x}_{DM}$, relative difference in sample means; $s_{DM}$, standard deviation of the difference in sample means; $rs_{DM}$, relative standard deviation of the difference in sample means; BF, Bayes Factor; $p_{NHST}$, p-values from null hypothesis significance test; $p_E$, p-value from two one sided t-tests; $p_\delta$, p-value from two one sided t-tests; CD, cohen's d; $\delta_M$, most difference in means; $r\delta_M$, relative most mean difference in sample means; Rnd, random.*

**Table S2:** Loss functions for Each Measure of Null strength

| Measure | Loss Functions: $Loss(x, y, x', y', \theta, \theta') :=$ | Eq. |
|---|---|---|
| $|\mu_{DM}|$: | $1 - \mathbb{I}(|\mu_{DM}| < |\mu'_{DM}| \text{ and } |\delta(x, y, \alpha_{DM})| > |\delta(x', y', \alpha'_{DM})|)$ | *(S8)* |
| $\sigma_D$: | $1 - \mathbb{I}(\sigma_D < \sigma'_D \text{ and } |\delta(x, y, \alpha_{DM})| > |\delta(x', y', \alpha'_{DM})|)$ | *(S9)* |
| $df_D$: | $1 - \mathbb{I}(df_D > df'_D \text{ and } |\delta(x, y, \alpha_{DM})| > |\delta(x', y', \alpha'_{DM})|)$ | *(S10)* |
| $\alpha_{DM}$: | $1 - \mathbb{I}(\alpha_{DM} > \alpha'_{DM} \text{ and } |\delta(x, y, \alpha_{DM})| > |\delta(x', y', \alpha'_{DM})|)$ | *(S11)* |
| $|r\mu_{DM}|$: | $1 - \mathbb{I}(|r\mu_{DM}| < |r\mu'_{DM}| \text{ and } |\delta(x, y, \alpha_{DM})| > |\delta(x', y', \alpha'_{DM})|)$ | *(S12)* |
| $r\sigma_D$: | $1 - \mathbb{I}(r\sigma_{DM} < r\sigma'_{DM} \text{ and } |\delta(x, y, \alpha_{DM})| > |\delta(x', y', \alpha'_{DM})|)$ | *(S13)* |

*Note: decision rule for candidate prediction ($\delta$) may by a "greater than" or "less than" operation depending on candidate statistic. The loss functions specifies when the prediction disagrees with the ground truth designation. In this case, the ground truth designations are for higher null strength for experiment 1 vs experiment 2 ('), and the candidate predictions test for lower null strength for experiment 1.*

**Table S3:** Null Results for Total Plasma Cholesterol

| | r$\bar{x}_{DM}$ | Group X | $\bar{x}$ | $s_X$ | m | Group Y | $\bar{y}$ | $s_Y$ | n | Units | $\alpha_{DM}$ | Sp | PMID, Loc | NE |
|---|---|---|---|---|---|---|---|---|---|---|---|---|---|---|
| 1 | -6% | Alfp-cre | 3.45 | 0.24 | 6 | Alfp-creTRβ$^{fl/fl}$ | 3.26 | 0.22 | 6 | mmol/L | 0.05 | ms | 24797634, F1C | ✓ |
| 2 | -6% | ApoE$^{-/-}$CD6$^{WT}$♂ | 1251 | 161 | 10 | ApoE$^{-/-}$CD6$^{-/-}$ ♂ | 1179 | 143 | 5 | mg/dl | 0.05 | ms | 29615096, T1 | ✓ |
| 3 | 0% | Cyp17A$^{WT}$♂ | 2.29 | 0.53 | 7 | Cyp17A$^{-/-}$♂ | 2.3 | 0.32 | 6 | mmol/L | 0.05 | ms | 32472014, T2 | |
| 4 | 7% | VLDLr0$^{-/-}$LpL1$^{WT}$ | 141 | 34 | 8 | VLDLr0$^{-/-}$LpL1$^{-/-}$ | 151 | 24 | 13 | mg/dl | 0.05 | ms | 11790777, T1 | |
| 5 | -11% | LDLr$^{-/-}$ | 105 | 28 | 15 | LDLr$^{-/-}$ + IFD | 93.4 | 29 | 15 | mmol/L | 0.05/3 | ms | 9614153, T2 | ✓ |
| 6 | 14% | ApoE$^{-/-}$CD6$^{WT}$♂ | 2518 | 257 | 8 | ApoE$^{-/-}$CD6$^{-/-}$♂ | 2876 | 506 | 6 | mg/dl | 0.05 | ms | 29615096, T1 | ✓ |
| 7 | -2% | Vehicle | 1335 | 269 | 8 | Probucol | 1303 | 376 | 8 | mg/dL | 0.05/12 | rb | 24188322, F1 | ✓ |
| 8 | 31% | ApoE$^{-/-}$ | 568 | 81 | 6 | ApoE$^{-/-}$ + PAO | 742 | 256 | 4 | mg/dL | 0.05 | ms | 27683551, F5E | |
| 9 | 37% | Control | 11.2 | 8.78 | 8 | Probucol | 15.3 | 6.6 | 8 | mg/dL | 0.05 | ms | 8040256, T1 | |

*Abbreviations: NE, negligible effect; $\bar{x}$, $\bar{y}$, sample means of group X and Y; $s_X$, $s_Y$, sample standard deviations of group X and Y; Sp, species; PMID, PubMed ID; Loc, location in manuscript; ms, mouse; rb, rabbit; pg, pig; mc, macaque; mk, monkey; hu, human (see respective publications for abbreviations used in group names).*

**Table S4:** Positive Results for Total Cholesterol

| r$\bar{x}_{DM}$ | Group X | $\bar{x}$ | $s_X$ | m | Group Y | $\bar{y}$ | $s_Y$ | n | Units | $\alpha_{DM}$ | Sp | PMID, Loc |
|---|---|---|---|---|---|---|---|---|---|---|---|---|
| -21% | ApoE$^{-/-}$ PAI-1$^{WT}$ | 2503 | 266 | 11 | ApoE$^{-/-}$ PAI-1$^{-/-}$ | 1984 | 252 | 13 | mg/dl | 0.05 | ms | 10712412, T1 |
| -29% | Ldlr$^{-/-}$Ad-Gal-Reln$^{FL/FL}$ | 2087 | 531 | 15 | Ldlr$^{-/-}$ Ad-Cre-Reln$^{FL/FL}$ | 1487 | 364 | 16 | mg/dl | 0.05 | ms | 26980442, SF2B |
| -30% | Vehicle | 1335 | 269 | 8 | Atorvastin | 934 | 232 | 8 | mg/dL | 0.05/12 | rb | 24188322, F1 |
| -31% | WTD -IF | 4.78 | 1.21 | 20 | WTD +IF | 3.3 | 0.67 | 20 | mmol/L | 0.05/4 | ms | 9614153, F1A |
| -33% | Placebo | 202 | 28.2 | 5 | 150 mg Atorvastatin | 135.2 | 64.2 | 5 | mg/dL | 0.05/6 | hu | 22716983, ST9 |
| -36% | Luciferase siSRNA | 238 | 15.7 | 5 | Angptl3 siRNA | 153 | 22.4 | 5 | mg/dL | 0.05/6 | ms | 32808882, F1D |
| -45% | WTD +Saline | 463 | 103 | 12 | WTD +PCSK9-mAb1 | 254 | 108 | 10 | mg/dl | 0.05 | ms | 31366894, F1A |
| -56% | pCMV5 | 642 | 63 | 9 | PCMV-E3 | 283 | 69 | 10 | mg/dl | 0.05 | ms | 11110410, F3 |
| -58% | ApoE$^{-/-}$ | 300 | 97 | 14 | ApoE$^{-/-}$ +Palm-E | 126 | 41 | 14 | mg/dl | 0.05 | ms | 11015467, T1 |

*Note: for abbreviations see Table S3.*

**Table S5:** Null Results for Plaque Area

| | r$\bar{x}_{DM}$ | Group X | $\bar{x}$ | s$_X$ | m | Group Y | $\bar{y}$ | s$_Y$ | n | Units | α$_{DM}$ | Sp | PMID, Loc | NE |
|---|---|---|---|---|---|---|---|---|---|---|---|---|---|---|
| 1 | 6% | Saline | 48.9 | 3.3 | 9 | LNA-Control | 51.6 | 5.2 | 13 | % | 0.05/3 | ms | 27137489, F1C | ✓ |
| 2 | 1% | ApoE$^{-/-}$ | 668500 | 1E+05 | 8 | ApoE$^{-/-}$.Yaa | 673300 | 2E+05 | 8 | μm2 | 0.05 | ms | 33110193, F1A | |
| 3 | -2% | Vehicle | 365693 | 1E+05 | 14 | PF-'2999 | 358500 | 2E+05 | 15 | μm² | 0.05/1 | ms | 30889221, F2B | ✓ |
| 4 | -18% | LDLr$^{-/-}$-IF | 360 | 94 | 9 | LDLr$^{-/-}$+IF | 295 | 36.0 | 8 | mm² | 0.05/1 | ms | 9614153, F2A | ✓ |
| 5 | -16% | LDLr$^{-/-}$ Fxr$^{WT}$ ♀ | 23.1 | 11.3 | 7 | LDLr$^{-/-}$ Fxr$^{-/-}$ ♀ | 19.4 | 7.2 | 8 | % | 0.05/2 | ms | 16825595, F2B | ✓ |
| 6 | 14% | Prox. -Sten. | 36 | 23 | 15 | Prox. +Sten. | 41 | 50 | 13 | % | 0.05 | mk | 3795393, T3 | ✓ |
| 7 | -13% | Control | 0.713 | 0.297 | 8 | BMY22089 | 0.617 | 0.45 | 8 | mm² | 0.05/7 | rb | 7840808, T4 | ✓ |
| 8 | 2% | Vehicle | 304788 | 1E+05 | 4 | SC-64258 | 310284 | 2E+05 | 3 | μm² | 0.05/3 | pg | 10571535, T2 | ✓ |
| 9 | 73% | ApoE$^{-/-}$-Abx | 9956 | 11578 | 20 | ApoE$^{-/-}$+Abx | 17196 | 13373 | 18 | μm² | 0.05/6 | ms | 21475195, F5E | ✓ |

*Note: for abbreviations see Table S3.*

**Table S6:** Positive Results for Plaque Area

| r$\bar{x}_{DM}$ | Group X | $\bar{x}$ | s$_X$ | m | Group Y | $\bar{y}$ | s$_Y$ | n | Units | α$_{DM}$ | Sp | PMID, Loc |
|---|---|---|---|---|---|---|---|---|---|---|---|---|
| -24% | ApoE$^{-/-}$ | 0.54 | 0.12 | 16 | ApoE$^{-/-}$P2Y$_1$$^{-/-}$ | 0.41 | 0.116 | 15 | mm² | 0.05 | ms | 18663083, F2B |
| -28% | ApoE$^{-/-}$ | 225000 | 72732 | 10 | ApoE$^{-/-}$+ PAO | 161000 | 47434 | 10 | μm² | 0.05 | ms | 27683551, F5C |
| -42% | ApoE$^{-/-}$ | 21.2 | 8.083 | 6 | ApoE$^{-/-}$P2Y$_1$$^{-/-}$ | 12.4 | 2.939 | 6 | % | 0.05 | ms | 18663083, F1A |
| -47% | ApoE$^{-/-}$ | 16.1 | 7.7 | 10 | ApoE$^{-/-}$EC-TFEB | 8.58 | 3.3 | 12 | % | 0.05 | ms | 28143903, F7F |
| -51% | Vehicle | 304788 | 1E5 | 4 | SC-69000 | 149779 | 34576 | 7 | μm² | 0.05/3 | pg | 10571535, T2 |
| -52% | ApoE$^{-/-}$ | 46.2 | 10.6 | 3 | ApoE$^{WT}$ | 22.1 | 8 | 3 | % | 0.05 | pg | 30305304, F5A |
| -69% | Progression Cotrol | 0.713 | 0.297 | 8 | Atorvastatin | 0.221 | 0.147 | 8 | mm² | 0.05/7 | rb | 7840808, T4 |
| -73% | LDLr$^{-/-}$-IF | 7.39 | 6.7 | 13 | LDLr$^{-/-}$+IF | 2.01 | 3.4 | 15 | % | 0.05 | ms | 9614153, F2B |
| -91% | Vehicle | 0.173 | 0.150 | 8 | Probucol | 0.015 | 0.025 | 8 | mm² | 0.05/6 | rb | 24188322, F1 |

*Note: for abbreviations see Table S3.*

**Supplementary References**